\documentclass[11pt,twoside]{article}
\usepackage{jeffe,fullpage,graphicx}
\usepackage[tilde,same]{myurl}

\newtheorem{lemma}{Lemma}[section]
\newtheorem{theorem}[lemma]{Theorem}
\newtheorem{corollary}[lemma]{Corollary}

\newtheorem{claim}{Claim}[lemma]

\urldef{\paperURL}\url{http://www.cs.uiuc.edu/~jeffe/pubs/screw.html}
\graphicspath{:Fig}

\def\Real{\mathbb{R}}
\def\Complex{\mathbb{C}}
\def\Sphere{\mathbb{S}}
\def\Hyper{\mathbb{S}}
\def\x#1{\widehat{\smash{#1}\vphantom{R}}}

\begin{document}

\pagestyle{myheadings}
\markboth{Jeff Erickson}
	 {Dense Point Sets Have Sparse Delaunay Triangulations}

\begin{titlepage}

\title{Dense Point Sets Have Sparse Delaunay Triangulations%
	\thanks{Portions of this work were done while the author was
	visiting The Ohio State University.  This research was
	partially supported by an Sloan Research Fellowship, by NSF
	CAREER grant CCR-0093348, and by NSF ITR grants DMR-0121695
	and CCR-0219594.  An extended abstract of this paper was
	presented at the 13th Annual ACM-SIAM Symposium on Discrete
	Algorithms \cite{e-dpshs-02}.  See \paperURL\ for the most
	recent version of this paper.}
	\\[0.5ex]
	\large or ``\dots But Not Too Nasty''%
}

\author{Jeff Erickson\\[1ex]
	\normalsize
	\begin{tabular}{c}
	University of Illinois, Urbana-Champaign\\
	jeffe@cs.uiuc.edu\\
	\url{http://www.cs.uiuc.edu/~jeffe}
	\end{tabular}
}

\date{Submitted to \textsl{Discrete \& Computational Geometry} --- \today}

\maketitle

\begin{bigabstract}
The \emph{spread} of a finite set of points is the ratio between the
longest and shortest pairwise distances.  We prove that the Delaunay
triangulation of any set of $n$ points in~$\Real^3$ with spread
$\Delta$ has complexity $O(\Delta^3)$.  This bound is tight in the
worst case for all $\Delta = O(\sqrt{n})$.  In particular, the
Delaunay triangulation of any dense point set has linear complexity.
We also generalize this upper bound to regular triangulations of
$k$-ply systems of balls, unions of several dense point sets, and
uniform samples of smooth surfaces.  On the other hand, for any $n$
and $\Delta = O(n)$, we construct a regular triangulation of
complexity $\Omega(n\Delta)$ whose $n$ vertices have spread $\Delta$.
\end{bigabstract}

\thispagestyle{empty}
\setcounter{page}{0}
\end{titlepage}


\section{Introduction}

Delaunay triangulations and Voronoi diagrams are one of the most
thoroughly studied objects in computational geometry, with
applications to nearest-neighbor searching \cite{ar-ddsnn-91,
c-papop-85, dv-cprac-77, gks-ricdv-92}, clustering
\cite{aesw-emstb-91, iki-awvdr-94, kl-chcem-95, s-vdbak-91},
finite-element mesh generation \cite{cdeft-se-99, elmsttuw-scus-00,
lt-gwsdm-01, s-tmgdr-98}, deformable surface modeling
\cite{cdes-dst-01}, and surface reconstruction \cite{ab-srvf-99,
abk-nvbsr-98, acdl-sahsr-00, ack-pcubm-01, bc-ssrnn-00, hs-vbihc-00}.
Many algorithms in these application domains begin by constructing the
Delaunay triangulation or Voronoi diagram of a set of points in
$\Real^3$.  Since three-dimensional Delaunay triangulations can have
complexity $\Omega(n^2)$ in the worst case, these algorithms have
worst-case running time $\Omega(n^2)$.  However, this behavior is
almost never observed in practice except for highly-contrived inputs.
For all practical purposes, three-dimensional Delaunay triangulations
appear to have linear complexity.

This frustrating discrepancy between theory and practice motivates our
investigation of practical geometric constraints that imply
low-complexity Delaunay triangulations.  Previous research on this
topic has focused on \emph{random} point sets under various
probability distributions \cite{b-ec3dv-90, m-iaeln-53, g-rssc-62,
m-rds-72, d-hdvdl-91, d-enkfv-93, e-npsch-01, gn-ac3dv-00,
gn-pcvdp-02}; \emph{well-spaced} point sets, which are low-discrepancy
samples of Lipschitz density functions \cite{cdeft-se-99, lt-gwsdm-01,
mtg-ocum-99, mttw-dbnmt-95, t-wspnm-97, t-pssug-92}; and \emph{surface
samples} with various density constraints \cite{ab-cdtpp-01,
ab-lbcdt-02, e-npsch-01, gn-ac3dv-00, gn-pcvdp-02}.  (We will discuss
the connections between these models and our results in
Section~\ref{S:related}.)  Our efforts fall under the rubric of
\emph{realistic input models}, which have been primarily studied for
inputs consisting of polygons or polyhedra \cite{bksv-rimga-97,
v-ffrim-97,z-ssgc-00} or sets of balls \cite{ho-smhsr-98,z-ssgc-00}.

This paper investigates the complexity of three-dimensional Delaunay
triangulations in terms of a geometric parameter called the
\emph{spread}, continuing our work in an earlier paper
\cite{e-npsch-01}.  The spread of a set of points is the ratio between
the largest and smallest interpoint distances.  Of particular interest
are \emph{dense} point sets in $\Real^d$, which have spread
$O(n^{1/d})$.  Valtr and others \cite{akp-mscpr-89, evw-cdpsh-97,
v-cis7h-92, v-ppsbr-94, v-llpic-96, v-acpdp-97} have established
several combinatorial results for dense point sets that improve
corresponding bounds for arbitrary point sets.  For example, a dense
point set in $\Real^3$ has at most $O(n^{7/3})$ halving planes; the
best upper bound known for arbitrary point sets is $O(n^{5/2})$
\cite{sst-ibkst-00}.  For other combinatorial and algorithmic results
related to spread, see \cite{a-gmsid-90, cs-pmsps-98, c-nnqms-99,
gimv-gpmps-99, gps-crotr-89, gps-iscir-90, imv-gmncb-99}.

In Section~\ref{S:del}, we prove that the Delaunay triangulation of
any set of points in~$\Real^3$ with spread~$\Delta$ has complexity
$O(\Delta^3)$.  This upper bound is independent of the number of
points in the set.  In particular, the Delaunay triangulation of any
dense point set in $\Real^3$ has only linear complexity.  This bound
is tight in the worst case for all $\Delta = O(\sqrt{n})$ and improves
our earlier upper bound of $O(\Delta^4)$~\cite{e-npsch-01}.

Our upper bound can be extended in several ways.  To make the notion
of spread less sensitive to close pairs, we define the \emph{order-$k$
spread} $\Delta_k$ to be the ratio of the diameter of the set to the
radius of the smallest ball containing $k$ points.  Our proof almost
immediately implies that the Delaunay triangulation has complexity
$O(k^2\Delta_k^3)$ for any $k$.  Our techniques also generalize fairly
easily to regular triangulations of disjoint balls whose centers have
spread $\Delta$.  With somewhat more effort, we show that if a set of
points can be decomposed into $k$ subsets, each with spread $\Delta$,
then its Delaunay triangulation has spread $O(k^2\Delta^3)$.  Our
results also imply upper bounds on the complexity of the Delaunay
triangulation of uniform or random samples of surfaces.  Finally, our
combinatorial bounds imply that the standard randomized incremental
algorithm~\cite{gks-ricdv-92} constructs the Delaunay triangulation
any set of points in expected time $O(\Delta^3\log n)$.  These and
other related results are developed in Section~\ref{S:imps}.

However, our upper bound does not generalize to arbitrary
triangulations.  In Section \ref{S:nondel}, for any~$n$ and $\Delta
\le n$, we construct a regular triangulation, whose $n$ vertices have
spread~$\Delta$, whose overall complexity is $\Omega(n\Delta)$.  (The
defining balls for this triangulation overlap heavily.)  This
worst-case lower bound was already known for Delaunay triangulations
for all ${\sqrt{n} \le \Delta \le n}$ \cite{e-npsch-01}.  In
particular, there is a dense point set in $\Real^3$, arbitrarily close
to a cubical lattice, with a regular triangulation of complexity
$\Omega(n^{4/3})$.

\medskip Throughout the paper, we analyze the complexity of
three-dimensional Delaunay triangulations by counting their edges.
Since the link of every vertex in a three-dimensional triangulation is
a planar graph, Euler's formula implies that any triangulation with
$n$ vertices and $e$ edges has at most $2e-2n$ triangles and $e-n$
tetrahedra.  Two points are joined by an edge in the Delaunay
triangulation of a set~$S$ if and only if they lie on a sphere with no
points of~$S$ in its interior.


\section{Previous and related results}
\label{S:related}

\subsection{Points in Space}

\begin{figure}[htb]
\centering\footnotesize\sf
\begin{tabular}{c@{\qquad}c}
 \includegraphics[height=2.25in]{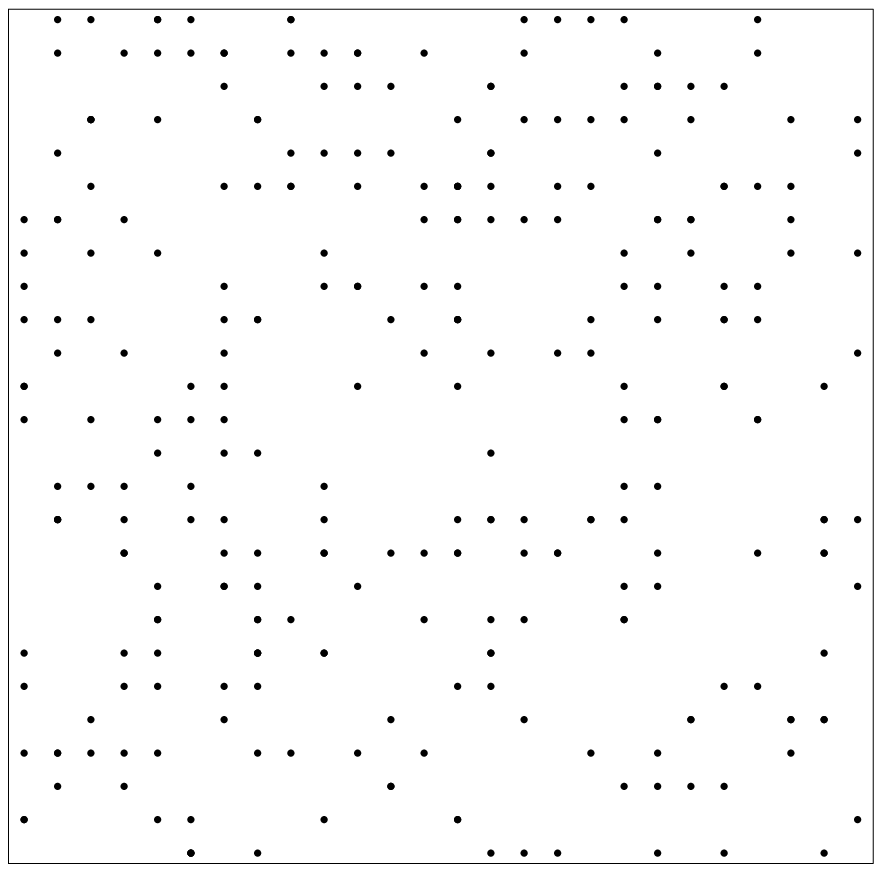} & 
 \includegraphics[height=2.25in]{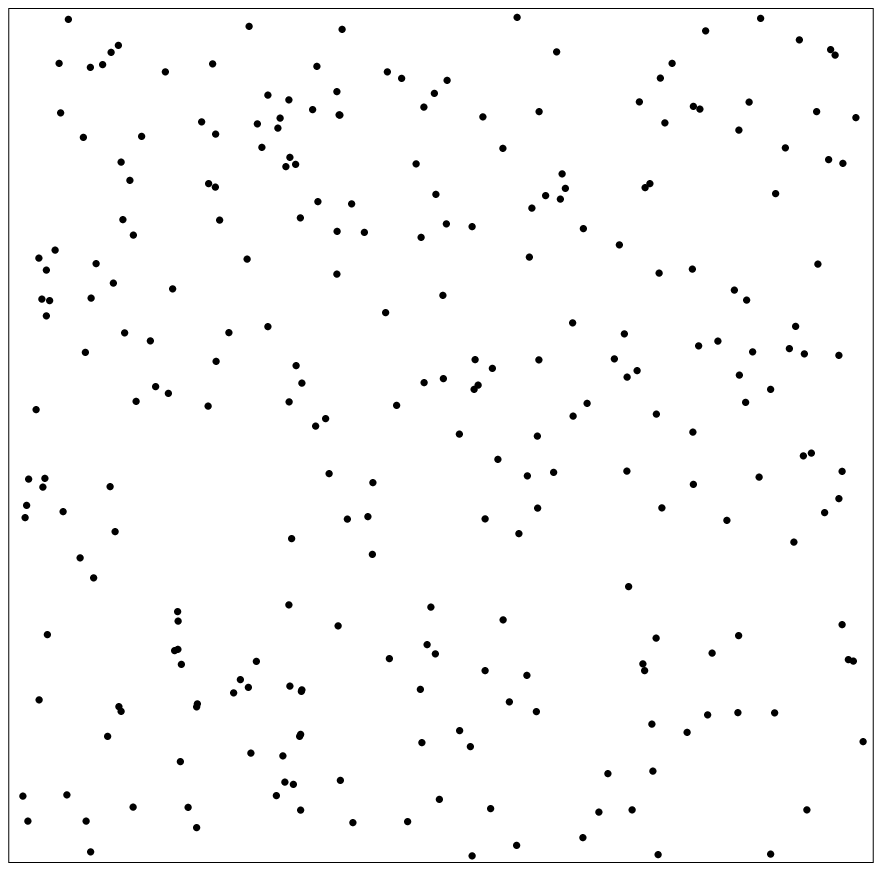}
\\
 (a) & (b)
\\[2ex]
 \includegraphics[height=2.25in]{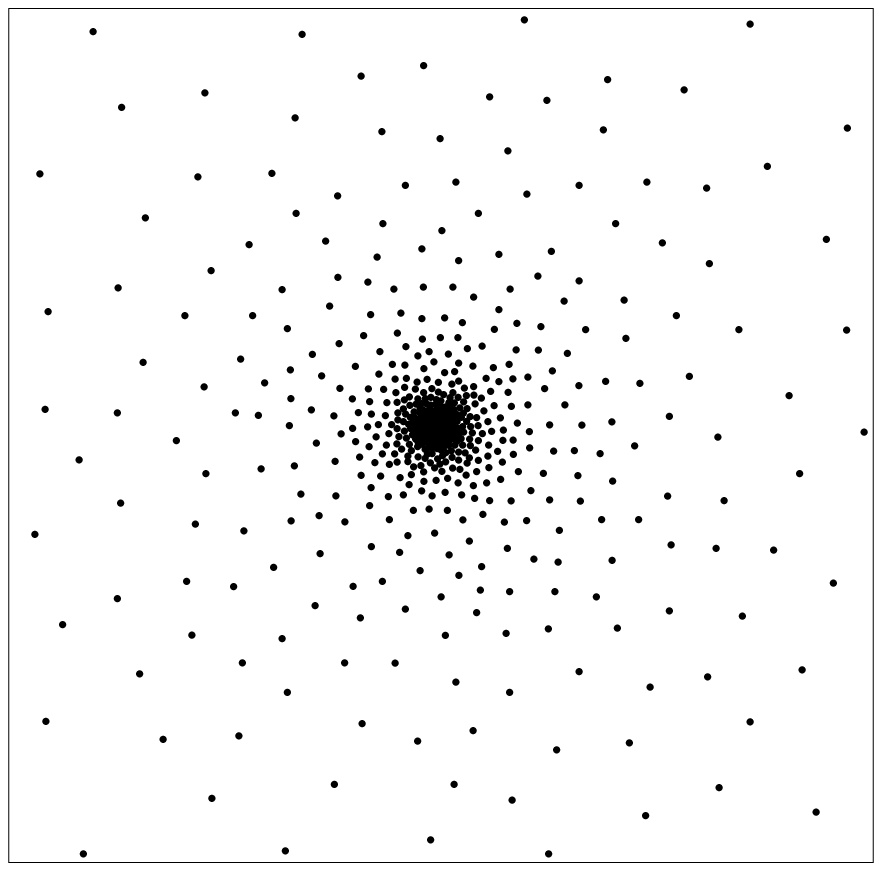} &
 \includegraphics[height=2.25in]{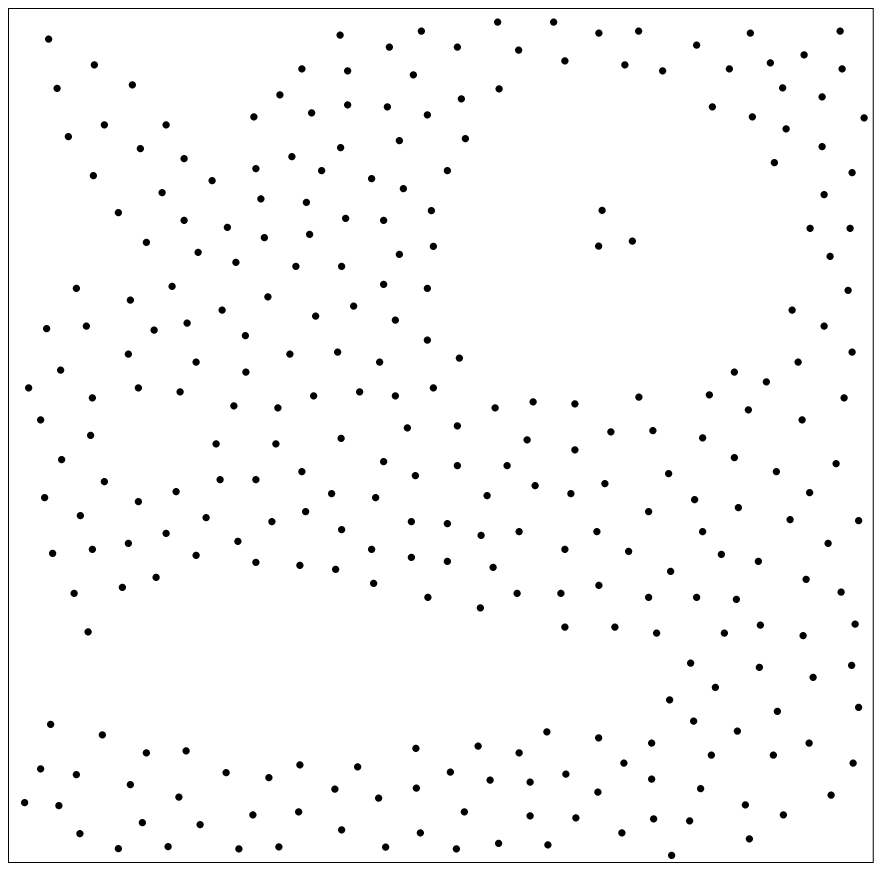}
\\
 (c) & (d)
\end{tabular}
\caption{Four models of `realistic' point sets. (a) Small
integer.  (b) Random.  (c) Well-spaced.  (d) Dense.}
\label{Fig/points}
\end{figure}

Our results for dense sets compare favorably with three other types of
`realistic' point data: points with small integer coordinates, random
points, and well-spaced points.  Figure \ref{Fig/points} illustrates
these four models.  Although the results for these models are quite
similar, we emphasize that with one exception---integer points with
small coordinates are dense---results in each model are formally
incomparable with results in any other model.

Unlike our new results, which apply only to points in $3$-space, all
of these related results have been generalized to higher dimensions.
We conjecture that the Delaunay triangulation of any $d$-dimensional
point set with spread $\Delta$ has complexity $O(\Delta^d)$, but new
proof techniques will be required to prove this bound.

\paragraph{Integer points.}
First, we easily observe that any triangulation of $n$ points in
$\Real^3$ with integer coordinates between $1$ and $\Delta$ has
complexity $O(\Delta^3)$, since each tetrahedron has volume at least
$1/6$.  (It is an open question whether this bound is tight for all
$n$ and $\Delta$.)  In particular, if $\Delta = O(n^{1/3})$, so that
the set is dense, the complexity of the Delaunay triangulation is
$O(n)$.  Dense sets obviously need not lie on a coarse integer grid;
nevertheless, this observation provides some useful intuition for our
results.

\paragraph{Random points.}
Statistical properties of Voronoi diagrams of random points have been
studied for decades, much longer than any systematic algorithmic
development.  In the early 1950s, Meijering \cite{m-iaeln-53} proved
that for a homogeneous Poisson process in $\Real^3$, the expected
number of Delaunay neighbors of any point is $48\pi^2/35 + 2 \approx
15.54$; see Miles~\cite{m-rds-72}.  This result immediately implies
that the Delaunay triangulation of a sufficiently dense random
periodic point set has linear expected complexity.  See M{\o}ller
\cite{m-rtm-89} for generalizations to arbitrary dimensions and Okabe
\etal~\cite[Chapter 5]{obsc-stcav-00} for an extensive survey of
statistical properties of random Voronoi diagrams and Delaunay
triangulations.

Bentley \etal~\cite{bwy-oetac-80} proved that for $n$ uniformly
distributed points in the $d$-dimensional hypercube, the expected
number of points with more than a constant number of Delaunay
neighbors is $O(n^{1-1/d}\log n)$; all such points lie near the
boundary of the cube.  Extending their technique, Bernal
\cite{b-ec3dv-90} proved that the expected complexity of the Delaunay
triangulation of $n$ random points in the three-dimensional cube is
$O(n)$.  Dwyer~\cite{d-hdvdl-91, d-enkfv-93} showed that if a set of
$n$ points is generated uniformly at random in the unit ball in
$\Real^d$, the Delaunay triangulation has expected complexity
$d^{O(d)}n$; in particular, each point has $d^{O(d)}$ Delaunay
neighbors on average.

Random point sets are not dense, even in expectation.  Let $S$ be a
set of $n$~points, generated independently and uniformly from the unit
hypercube in~$\Real^d$.  A straightforward `balls and bins'
argument~\cite{e-ptnna-67,mr-ra-95} implies that the expected spread
of $S$ is $\Theta(n^{2/d})$; moreover, with high
probability,\footnote{that is, with probability $1-n^{-c}$ for some
constant $c>0$} the spread of $S$ is between $\Omega(n^{2/d}/\log^\e
n)$ and $O(n^{2/d+\e})$ for any~$\e$.

\paragraph{Well-spaced points.}
Miller, Talmor, Teng and others~\cite{mtg-ocum-99, mttw-dbnmt-95,
t-wspnm-97, t-pssug-92} have derived several results for
\emph{well-spaced} point sets in the context of high-quality mesh
generation.  A point set $S$ in $\Real^3$ is well-spaced with respect
to a $1$-Lipschitz spacing function $\lambda:\Real^3\to\Real^+$ if,
for some fixed constants $0 < \delta < 1/2$ and $0 < \e < 1$, the
distance from any point $x\in\Real^3$ to its second nearest
neighbor\footnote{If $x$ is a point in $S$, then $x$ is its own
nearest neighbor in $S$.  This is not the definition actually proposed
by Miller \etal, but it is easy to prove that our definition is
equivalent to theirs.} $p\in S$ is between $\delta\e\lambda(p)$ and
$\e\lambda(p)$.  In her thesis, Talmor \cite{t-wspnm-97} proves that
Delaunay triangulations of well-spaced point sets have complexity
$O(n)$; in particular, any point in a well-spaced point set has $O(1)$
Delaunay neighbors.

Any point set that is well-spaced with respect to a constant spacing
function is dense.  In general, however, the spread of a well-spaced
set can be exponentially large; consider the one-dimensional
well-spaced set $\set{2^{-i}\mid 1\le i\le n}$.  On the other hand,
dense point sets can contain large gaps and thus are not necessarily
well-spaced with respect to \emph{any} Lipschitz spacing function;
compare Figures~\ref{Fig/points}(c) and \ref{Fig/points}(d).

Talmor's linear upper bound \cite{t-wspnm-97} depends exponentially on
the spacing parameters $\delta$ and~$\e$, but this is largely an
artifact of the generality of her results.\footnote{Specifically,
Talmor first proves that the Delaunay triangulation of any well-spaced
point sets in any fixed dimensions is \emph{well-shaped}: for every
simplex, the ratio of circumradius to shortest edge length is bounded
by a constant.  She then proves that any vertex in any well-shaped
triangulation is incident to only a constant number of simplices.}
For the three-dimensional case, we can derive tighter bounds as
follows.  Let $S$ be a point set in $\Real^3$ that is well-spaced with
respect to some $1$-Lipschitz spacing function $\lambda$.  Let $p$ be
any point in $S$, and let $q$ and $r$ be any two Delaunay neighbors
of~$p$.  We immediately have the inequalities $\abs{pq} \le
\min\set{\e\lambda(p), \e\lambda(q)}$ and $\abs{qr} \ge
\delta\e\lambda(q)$.  Because the spacing function $\lambda$ is
Lipschitz, we have $\lambda(q) \ge \lambda(p) - \abs{pq} \ge
\lambda(p) - \e\lambda(q)$, which implies that $\lambda(q) \ge
\lambda(p)/(1+\e)$.  Together, these inequalities imply that the set
of Delaunay neighbors of $p$ has spread at most $2(1+\e)/\delta =
O(1/\delta)$.  A packing argument in our earlier
paper~\cite{e-npsch-01} now implies that $p$ has $O(1/\delta^2)$
Delaunay neighbors.  It follows that the Delaunay triangulation of $S$
has complexity $O(n/\delta^2)$.  (Surprisingly, this bound does not
depend on $\e$ at all!)

Finally, in contrast to both random and well-spaced point sets, a
single point in a dense set in~$\Real^3$ can have $\Theta(\Delta^2) =
\Theta(n^{2/3})$ Delaunay neighbors in the worst case
\cite{e-npsch-01}.  Thus, our upper bound proof must consider global
properties of the Delaunay triangulation.

\subsection{Points on Surfaces}
\label{SS:oldsurf}

The complexity of Delaunay triangulations of points on two-dimensional
surfaces in space has also been studied, largely due to the recent
proliferation of Delaunay-based surface reconstruction
algorithms~\cite{ab-srvf-99, abk-nvbsr-98, acdl-sahsr-00,
ack-pcubm-01, bc-ssrnn-00, hs-vbihc-00}.  Upper and lower bounds range
from linear to quadratic, depending on exactly how the problem is
formulated.  Specifically, the results depend on whether the surface
is considered fixed or variable, whether the surface is smooth or
polyhedral, and on the precise sampling conditions to be analyzed.  We
first review some standard terminology.

Let $\Sigma$ be a $C^2$ surface embedded in $\Real^3$.  A \emph{medial
ball} of $\Sigma$ is a ball whose interior is disjoint from $\Sigma$
and whose boundary touches $\Sigma$ at more than one point.  The
center of a medial ball is called a \emph{medial point}, and the
closure of the set of medial points is the \emph{medial axis} of
$\Sigma$.  The \emph{local feature size} of a point $x\in\Sigma$,
denoted $\lfs(x)$, is the the distance from $x$ to the medial axis.
Finally, a set of points $P\subset\Sigma$ is an \emph{$\e$-sample} of
$\Sigma$ if the distance from any surface point $x\in\Sigma$ to the
nearest sample point in $P$ is at most $\e\cdot\lfs(x)$.  This
condition imposes a lower bound on the number of sample points in any
region of the surface.  Given an $\e$-sample of an unknown
surface~$\Sigma$, for sufficiently small $\e$, the algorithms cited
above provably reconstruct a surface geometrically close and
topologically equivalent to $\Sigma$.  As a first step, each algorithm
constructs the Voronoi diagram of the sample points; the complexity of
this Voronoi diagram is clearly a lower bound on the running time of
the algorithm.

Unfortunately, $\e$-samples can have arbitrarily complex Delaunay
triangulations due to oversampling.  Specifically, for any surface
other than the sphere and any sampling density $\e>0$, there is an
$\e$-sample whose Delaunay triangulation has complexity $\Theta(n^2)$,
where $n$ is the number of sample points \cite{e-npsch-01}.  Thus, in
order to obtain nontrivial upper bounds, we must also impose an upper
bound on the density of samples.

Dey, Funke, and Ramos~\cite{dfr-sralt-01, fr-ssrnl-02}
define\footnote{Again, this is not the definition proposed by Dey
\etal, but it is easy to show that the definitions are equivalent.} a
set of points $P\subset \Sigma$ to be a \emph{locally uniform} sample
of $\Sigma$ if $P$ is well-spaced (in the sense of Miller \etal\@)
with respect to some $1$-Lipschitz function $\lambda:\Sigma\to\Real^+$
such that $\lambda(x) \le \lfs(x)$ for all $x\in\Sigma$.  If $P$ is
well-spaced with respect to the local feature size function, which is
always $1$-Lipschitz, we call $P$ a \emph{uniform} sample of
$\Sigma$~\cite{e-npsch-01}.  Dey \etal~\cite{dfr-sralt-01} described
an algorithm to reconstruct a surface from a locally uniform sample in
$O(n\log n)$ time; Funke and Ramos later showed to how extract a
locally uniform $\e$-sample from an arbitrary $\e$-sample in $O(n\log
n)$ time.  Neither of these algorithms constructs the Delaunay
triangulation of the points.

In our earlier paper~\cite{e-npsch-01}, we derived lower bounds on the 
complexity of Delaunay triangulations of uniform samples in terms of 
the \emph{sample measure}
\[
	\mu(\Sigma) = \iint_\Sigma \frac{dx^2}{\lfs(x)^2}.
\]
Any uniform $\e$-sample of $\Sigma$ contains $\Theta(\e^2\mu(\Sigma))$ 
points.  There are smooth connected surfaces with sample measure $\mu$ 
for which the Delaunay triangulation of \emph{any} uniform $\e$-sample 
has complexity $\Omega(\mu^2/\log^2\mu)$.

More positive results can be obtained by considering the surface to be 
fixed and considering the asymptotic complexity of the Delaunay 
triangulation as the number of sample points tends to infinity.  In 
this context, hidden constants in the upper bounds depend on geometric 
parameters of the fixed surface, such as the number of facets or their 
maximum aspect ratio for polyhedral surfaces, or the minimum curvature 
radius or sample measure for curved surfaces.  Since the surface is 
fixed, all such parameters are considered constants.

Golin and Na \cite{gn-ac3dv-00, gn-ac3dv-01, gn-ptldi-01} proved that
if $n$ points are chosen uniformly at random on the surface of any
fixed three-dimensional convex polytope, the expected complexity of
their Delaunay triangulation is $O(n)$.  Using similar techniques,
they recently showed that a random sample of a fixed \emph{nonconvex}
polyhedron has Delaunay complexity $O(n\log^4 n)$ with high
probability~\cite{gn-pcvdp-02}.  In fact, their analysis applies to 
any fixed set of disjoint triangles in $\Real^3$.

Attali and Boissonnat \cite{ab-lbcdt-02} recently proved that the
Delaunay triangulation of any \emph{$(\e,\kappa)$-sample} of a fixed
polyhedral surface has complexity $O(\kappa^2 n)$, improving their
previous upper bound of $O(n^{7/4})$ (for constant
$\kappa$)~\cite{ab-cdtpp-01}.  A set of points is called an
$(\e,\kappa)$-sample of a surface $\Sigma$ if every ball of
radius~$\e$ whose center lies on $\Sigma$ contains at least one and at
most $\kappa$ points in $P$.\footnote{This definition ignores the
local feature size, which is necessary for polyhedral surfaces, since
the local feature size is zero at any sharp corner.  Moreover, for any
fixed smooth surface, the minimum local feature size is a constant, so
any $(\e, \kappa)$-sample is an $O(\e)$-sample according to our
earlier definition.}  A simple application of Chernoff bounds (see
Theorem \ref{Th:randsurf}) implies that a random sample of $n$ points
on a fixed surface is an $O(\e, O(\log n))$-sample with high
probability, where $\e = O(\sqrt{(\log n)/n})$.  Thus, Attali and
Boissonnat's result improves Golin and Na's high-probability bound for
random points to $O(n\log^2 n)$.

Our new upper bound has a similar corollary.  Informally, a uniform
sample of any fixed (not necessarily polyhedral, smooth, or convex)
surface has spread $O(\sqrt{n})$, so its Delaunay triangulation has
complexity $O(n^{3/2})$.  This bound is tight in the worst case; 
a right circular cylinder with constant height and radius has a 
uniform $(\e,1)$-sample with Delaunay complexity $\Omega(n^{3/2})$.
Similar arguments establish upper bounds of $O(\kappa^2 n^{3/2})$ for
$(\e,\kappa)$-samples and $O(n^{3/2}\log^{3/2} n)$, with high
probability, for random samples.  We describe these results more
formally in Section~\ref{S:imps}.


\section{Sparse Delaunay Triangulations}
\label{S:del}

In this section, we prove the main result of the paper.

\begin{theorem}
\label{Th:del3}
The Delaunay triangulation of any finite set of points in $\Real^3$
with spread $\Delta$ has complexity $O(\Delta^3)$.
\end{theorem}

\begin{figure}[htb]
\centerline{\includegraphics[height=1.5in]{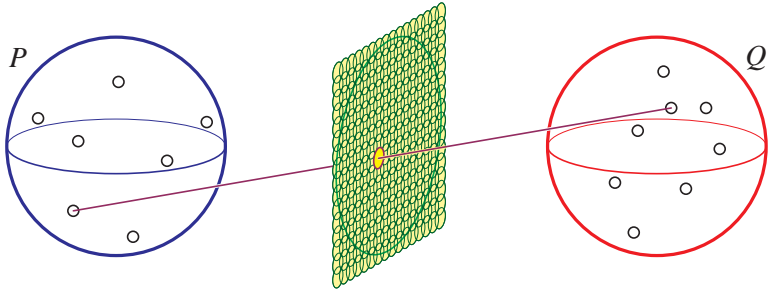}}
\caption{A well-separated pair of sets $P\cup Q$ and a crossing edge 
intersecting a pixel.}
\label{F:setup}
\end{figure}

Our proof is structured as follows.  We will implicitly assume that no
two points are closer than unit distance apart, so that spread is
synonymous with diameter.  Two sets $P$ and~$Q$ are
\emph{well-separated} if each set lies inside a ball of radius~$r$,
and these two balls are separated by distance~$2r$.  Without loss of
generality, we assume that the balls containing $P$~and~$Q$ are
centered at points $(0,0,2r)$ and $(0,0,-2r)$, respectively.  Our
argument ultimately reduces to counting the number of \emph{crossing
edges}---edges in the Delaunay triangulation of $P\cup Q$ with one
endpoint in each set.  See Figure~\ref{F:setup}.

Our proof has four major steps, each presented in its own subsection.

\begin{itemize}
\item
We place a grid of $O(r^2)$ circular \emph{pixels} of constant radius
$\e$ on the plane $z=0$, so that every crossing edge passes through a
pixel.  In Section \ref{SS:screw}, we prove that all the crossing
edges stabbing any single pixel lie within a slab of constant width
between two parallel planes.  Our proof relies on the fact that the
edges of a Delaunay triangulation have a consistent depth order from
any viewpoint~\cite{e-atccd-90}.

\item
We say that a crossing edge is \emph{relaxed} if its endpoints lie on
an empty sphere of radius $O(r)$.  In Section \ref{SS:slice}, we show
that at most $O(r)$ relaxed edges pass through any pixel, using a
generalization of the `Swiss cheese' packing argument used to prove
our earlier $O(\Delta^4)$ upper bound \cite{e-npsch-01}.  This implies
that there are $O(r^3)$ relaxed crossing edges overall.

\item
In Section~\ref{SS:twist}, we show that there are a constant number of 
conformal (\ie, sphere-preserving) transformations that change the 
spread of $P\cup Q$ by at most a constant factor, such that every 
crossing edge of $P\cup Q$ is a relaxed Delaunay edge in at least one 
image.  The proof uses a packing argument in a particular subspace of 
the space of three-dimensional \Mobius\ transformations.  It follows 
that $P\cup Q$ has at most $O(r^3)$ crossing edges.

\item
Finally, in Section \ref{SS:split}, we count the Delaunay edges for an
arbitrary point set $S$ using an octtree-based well-separated pair
decomposition \cite{ck-dmpsa-95}.  Every edge in the Delaunay
triangulation of $S$ is a crossing edge of some subset pair in the
decomposition.  However, not every crossing edge is a Delaunay edge; a
subset pair contributes a Delaunay edge only if it is close to a large
empty \emph{witness} ball.  We charge the pair's $O(r^3)$ crossing
edges to the $\Omega(r^3)$ volume of this ball.  We choose the witness
balls so that any unit of volume is charged at most a constant number
of times, implying the final $O(\Delta^3)$ bound.

\end{itemize}

\subsection{Nearly Concurrent Crossing Edges Are Nearly Coplanar}
\label{SS:screw}

The first step in our proof is to show that the crossing edges 
intersecting any pixel are nearly coplanar.  To do this, we use an 
important fact about \emph{depth orders} of Delaunay triangulations, 
related to shellings of convex polytopes.

Let $x$ be a point in $\Real^3$, called the \emph{viewpoint}, and
let~$S$ be a set of line segments (or other convex objects).  A
segment $s\in S$ is \emph{behind} another segment $t\in S$ with
respect to $x$ if $t$ intersects $\conv\set{x,s}$.  If the transitive
closure of this relation is a partial order, any linear extension is
called a \emph{consistent depth order} of $S$ with respect to $x$.
Otherwise, $S$ contains a \emph{depth cycle}---a sequence of segments
$s_1, s_2, \dots, s_k$ such that every segment $s_i$ is directly
behind its successor $s_{i+1}$ and $s_k$ is directly behind $s_1$.  De
Berg \etal~\cite{bos-cvdo-94} describe an algorithm that either
computes a depth order or finds a depth cycle for a given set of $n$
segments, in $O(n^{4/3+\e})$ time.  See de Berg \cite{b-rsdoh-93} and
Chazelle \etal~\cite{cegss-lsca-96} for related results.

\begin{figure}[htb]
\footnotesize\sf\centering
\begin{tabular}{c@{\qquad}c}
   \begin{tabular}{c}
      \includegraphics[height=0.75in]{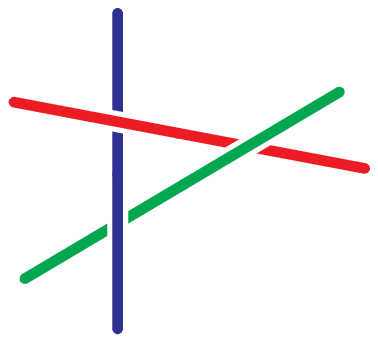} \\ (a) \\[2ex]
      \includegraphics[height=1.5in]{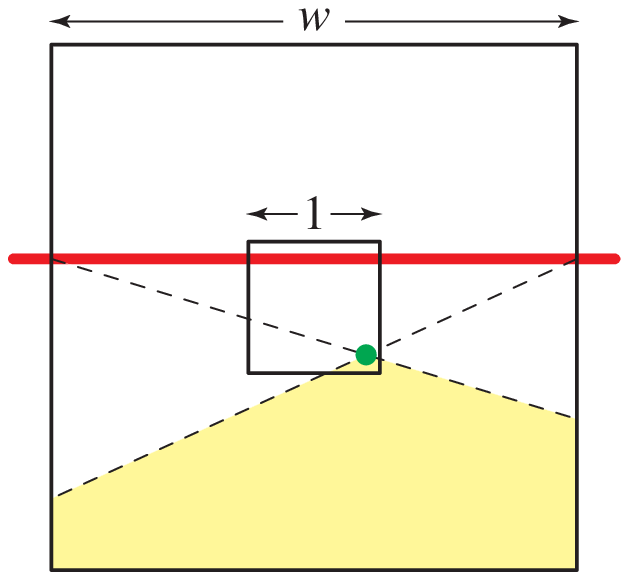} \\ (b)
   \end{tabular}
   &
   \begin{tabular}{c}
       \includegraphics[height=3in]{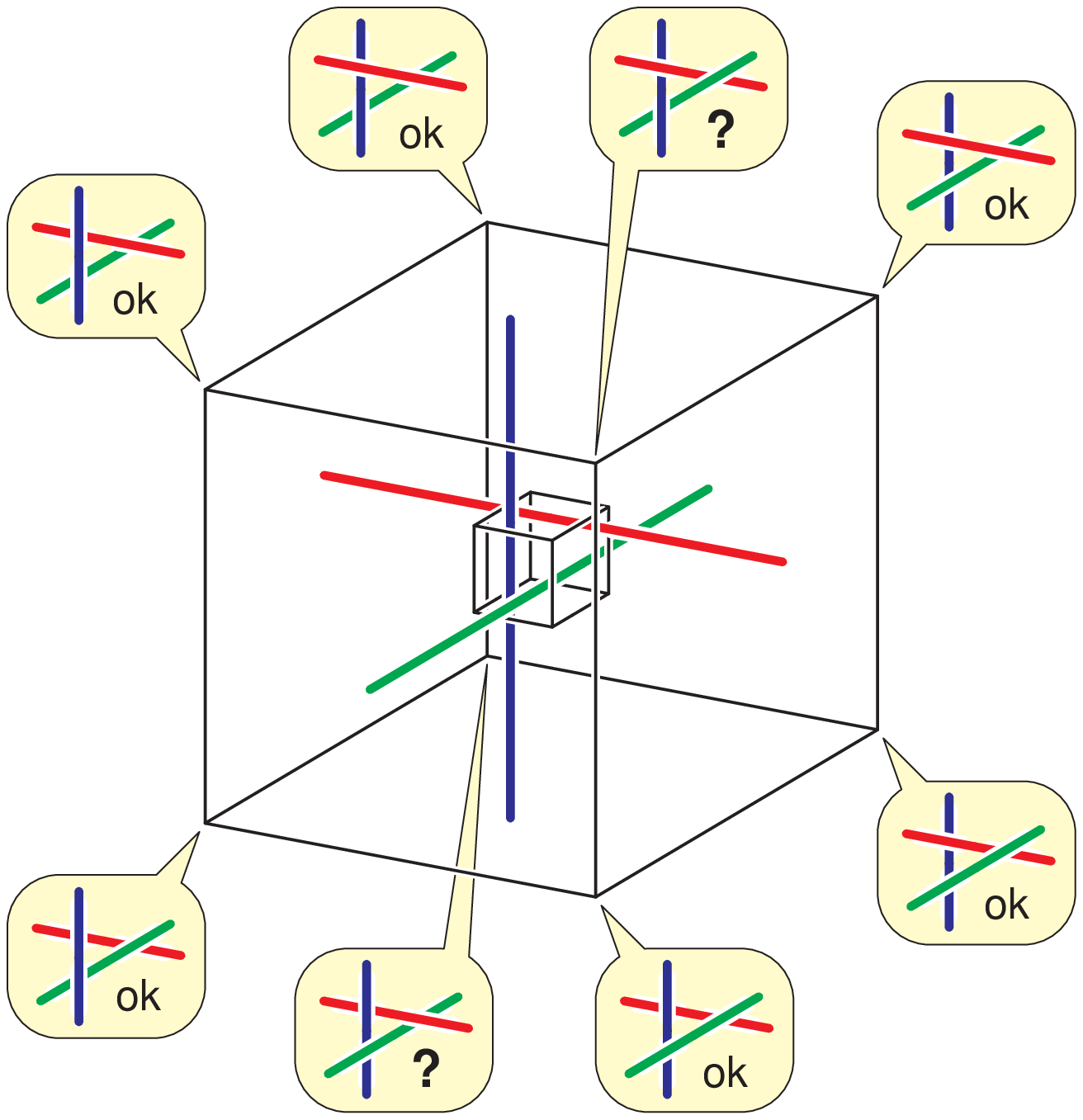} \\ (c)
   \end{tabular}
\end{tabular}
\caption{(a)~A screw.  (b)~Front view of $C$, showing viewpoints where
$s_x$ appears behind $s_z$.  (c)~Every vertex of $C$ sees a different
depth order, two of which are inconsistent.  See
Lemma~\protect\ref{L:screw}.}
\label{F:screw}
\end{figure}

We say that three line segments form a \emph{screw} if they form a
depth cycle from some viewpoint.  See Figure~\ref{F:screw}(a).

\begin{lemma}
\label{L:noscrew}
The edges of any Delaunay triangulation have a consistent depth order 
from any viewpoint.  In particular, no three Delaunay edges form a 
screw.
\end{lemma}

\begin{proof}
Let $x$ be a point, and let $S$ be a sphere with radius $r$ and center
$c$.  The \emph{power distance} from $x$ to $S$ is $\abs{xc}^2 - r^2$;
if $x$ is outside $S$, this is the square of the distance from $x$ to
$S$ along a line tangent to $S$.  Edelsbrunner \cite{e-atccd-90,
e-gtmg-01} proved that a consistent depth order for the simplices in
any Delaunay triangulation, with respect to any viewpoint $x$, can be
obtained by sorting the power distances from $x$ to the (empty)
circumspheres of the simplices.  (See Section \ref{SS:regular}.)  This
is precisely the order in which the Delaunay tetrahedra are computed
by Seidel's shelling convex hull algorithm \cite{s-chdch-86}.  We can
easily extract a consistent depth order for the Delaunay edges from
this simplex order.
\end{proof}

The next lemma describes sufficient (but not necessary) conditions for
three pairwise-skew segments to form a screw.

\begin{lemma}
\label{L:screw}
Let $c$ be a parallelepiped centered at the origin, and let $C =
w\cdot c$ for some $w\ge 2+\sqrt{5} \approx 4.2361$.  Three line
segments, each parallel to a different edge of $c$, form a screw if
they all intersect $c$ but none of their endpoints lie inside~$C$.
\end{lemma}

\begin{proof}
Since any affine image of a screw is also a screw, it suffices to
consider the case where $c$ and $C$ are concentric axis-aligned cubes
of widths $1$ and $w$, respectively.  Let $s_x, s_y, s_z$ be the three
segments, parallel to the $x$-, $y$-, and $z$-axes, respectively.  Any
ordered pair of these segments, say $(s_z, s_x)$, define an unbounded
polyhedral region $V_{xz}$ of viewpoints from which $s_x$ appears
behind~$s_z$.  The segment $s_z$ is the only bounded edge of $V_{xz}$,
and both of its endpoints are outside~$C$.  Thus, we can determine
which vertices of $C$ lie inside $V_{xz}$ by considering the
projection to the $xy$-plane.  From Figure \ref{F:screw}(b), we
observe that if $w \ge 2+\sqrt{5}$, then $V_{xz}$ contains exactly
half of the vertices of $C$, all with the same $y$-coordinate.  A
symmetric argument implies that the other four vertices lie
in~$V_{zx}$.  Similarly, $V_{xy}$ and $V_{yx}$ partition the vertices
of $C$ by their $z$-coordinates, and $V_{yz}$ and $V_{zy}$ partition
the vertices of $C$ along their $x$-coordinates.  Thus, each of the
eight vertices of $C$ sees one of the eight possible depth orders of
the three segments.  Since only six of these orders are consistent,
two vertices of $C$ see a depth cycle, implying that the segments form
a screw.  See Figure \ref{F:screw}(c).
\end{proof}

Recall that a pixel is a circle of radius $\e$ in the $xy$-plane.

\begin{lemma}
\label{L:slab}
The crossing edges passing through any pixel lie inside a slab of 
width $(20+9\sqrt{5})\e \approx 40.1246\e$ between two parallel 
planes.
\end{lemma}

\def\te{\tilde{e}}
\def\tE{\tilde{E}}

\begin{proof}
We will in fact prove a stronger statement.  Any two planes $h_1$ and
$h_2$ whose line of intersection lies in the $xy$-plane define an
\emph{anchored double wedge}, consisting of all the points above~$h_1$
and below~$h_2$ or vice versa.  We define the \emph{thickness} of an
anchored double wedge to be the width of the two-dimensional slab
obtained by intersecting the double wedge with the plane $z=r$.  We
claim that the crossing edges passing through any pixel lie inside an
$\e$-neighborhood of an anchored double wedge with thickness
$(6+3\sqrt{5})\e$.  The lemma follows immediately from this claim.

Without loss of generality, suppose the pixel $\pi$ is centered at the
origin, and let $E$ denote the set of crossing edges passing through
$\pi$.  Translate each edge in $E$ parallel to the $xy$-plane so that
it passes through the origin, and call the resulting set of segments
$\tE$.  We need to show that the segments $\tE$ lie in an anchored
double wedge of thickness $(6+3\sqrt{5})\e$.  Since all these segments
pass through the origin, it suffices to show that the intersection
points between $\tE$ and the plane $z=r$ lie in a two-dimensional
strip of width $(6+3\sqrt{5})\e$ between two parallel lines.  The
width of a set of planar points is determined by only three points, so
it suffices to check every triple of segments in~$\tE$.

Let $e_1, e_2, e_3$ be three arbitrary crossing edges in $E$, and let
$\te_1, \te_2, \te_3$ be the corresponding segments in $\tE$.  For
each $i$, let $p_i$ and $q_i$ be the intersection points of $\te_i$
with the planes $z=r$ and $z=-r$, respectively, and let $s_i \subseteq
\te_i$ be the segment between $p_i$ and $q_i$.  Finally, let $\omega$
be the width of the thinnest two-dimensional strip containing the
triangle $\triangle p_1 p_2 p_3$.  Observe that $\triangle$ contains a
circle of radius $\omega/3$.

Let $C$ denote the scaled Minkowski sum $(s_1+s_2+s_3)/3$; this is a
parallelepiped centered at the origin, with edges parallel to the
original crossing edges $e_i$.  Since each segment $s_i$ fits exactly
within the slab $-r \le z \le r$, so does the cuboid $C$.  The
intersection of $C$ with the $xy$-plane is the scaled Minkowski sum
$(\triangle p_1 p_2 p_3 + \triangle q_1 q_2 q_3)/2$; this hexagon also
contains a circle of radius $\omega/3$.

Let $c$ be the smallest parallelepiped similar to and concentric with
$C$ that contains the pixel $\pi$.  Since $\pi$ is a circle of radius
$\e$ in the $xy$-plane, $C$ is at least a factor of $\omega/3\e$
larger than $c$.  Lemma~\ref{L:noscrew} implies that the Delaunay
edges $e_1, e_2, e_3$ cannot form a screw.  Thus, by
Lemma~\ref{L:screw}, we must have $\omega/3\e < 2+\sqrt{5}$, or
equivalently, $\omega < (6+3\sqrt{5})\e$, as claimed.
\end{proof}

\subsection{Slabs Contain Few Relaxed Edges}
\label{SS:slice}

At this point, we would like to argue that any slab of constant width
contains only $O(r)$ crossing edges.  Unfortunately, this is not
true---a variant of our helix construction \cite{e-npsch-01} implies
that a slab can contain up to $\Omega(r^3)$ edges, $\Omega(r^2)$ of
which can pass through a single, arbitrarily small pixel.  However,
most of these Delaunay edges have extremely large empty circumspheres.

We say that a crossing edge is \emph{relaxed} if its endpoints lie on
the boundary of an empty ball with radius less than $4r$, and
\emph{tense} otherwise.  In this section, we show that few relaxed
edges pass through any pixel.  Once again, recall that that $\e$
denotes the pixel radius.

\begin{lemma}
\label{L:degree}
If $\e < 1/16$, then for any pixel $\pi$, each point in $P$ is an 
endpoint of at most one relaxed edge passing through $\pi$.
\end{lemma}

\begin{proof}
Suppose some point $p\in P$ is an endpoint of two crossing edges $pq$
and $pq'$, both passing through~$\pi$, where $\abs{pq} \ge \abs{pq'}$.
We immediately have $\angle qpq' \le 2 \tan^{-1} (\e/r)$ and
$\abs{qq'} \ge 1$.  See Figure \ref{Fig:onerelaxed}. Thus, the circle
through $p$,~$q$, and $q'$ has radius at least $1/(4\tan^{-1} (\e/r))
\approx r/4\e > 4r$.  Any empty circumsphere of $pq$ must have at
least this radius, so it must be tense.
\begin{figure}[htb]
\centering\includegraphics[height=1.25in]{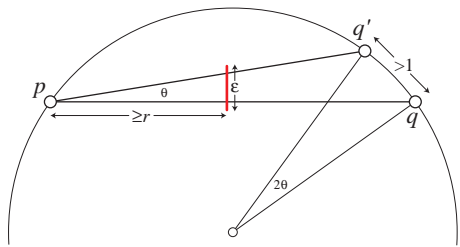}
\caption{Proof of Lemma \ref{L:degree}.  The radius of the circle is
at least $1/(4\tan^{-1}(\e/r))$, so edge $pq'$ is tense.}
\label{Fig:onerelaxed}
\end{figure}
\end{proof}

\begin{lemma}
\label{L:cheese}
The relaxed edges inside any slab of constant width are incident to at 
most $O(r)$ points in $P$.
\end{lemma}

\begin{proof}
Let $\sigma$ be a slab of width $\omega = O(1)$ between two parallel 
planes, and let $E$ be the set of points in $P$ incident to any 
relaxed edge contained in $\sigma$.  To prove that $\abs{E} = O(r)$, 
we use a variant of our earlier `Swiss cheese' packing 
argument~\cite{e-npsch-01}.  Intuitively, we take the intersection of 
the bounding sphere of $P$ and the slab~$\sigma$, remove the Delaunay 
circumspheres of relaxed edges in~$\sigma$, argue that the resulting 
`Swiss cheese slice' has small surface area, and then charge a 
constant amount of surface area to each endpoint in $E$.  To formalize 
this argument, we need to slightly expand $\sigma$ and slightly 
contract the Delaunay balls.

\begin{figure}[htb]
\centering\includegraphics[height=1.5in]{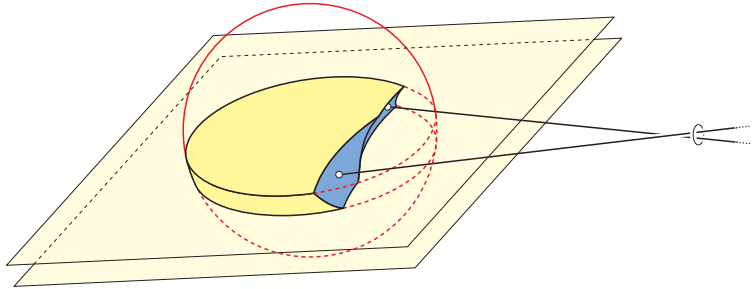}
\caption{The Swiss cheese slice $\Sigma$ determined by two relaxed
crossing edges intersecting a common pixel.  The darker (blue) portion
of the surface is $H$.  The contracted Delaunay balls $b_p$ are not
shown.}
\label{Fig:cheeseslice}
\end{figure}

Let $\sigma'$ be a parallel slab with the same central plane as
$\sigma$, with slightly larger width $\omega+1$.  Let $\bigcirc P$
denote the sphere of radius $r$ containing $P$.  Let $D$ be the
intersection of $\sigma'$ with the sphere of radius $r+1$ concentric
with $\bigcirc P$.  The volume of $D$ is at most $\pi(\omega+1)(r+1)^2
= O(r^2)$.

For each point $p\in E$, we define two balls: $B_p$ is the smallest
Delaunay ball of some relaxed edge $pq$, and $b_p$ is the open ball
concentric with $B_p$ but with radius smaller by $1/3$.  The radius
of~$B_p$ is at most~$4r$, and the radius of $b_r$ is at most $4r-1/3$.

Finally, we define the `Swiss cheese slice' $\Sigma = D \setminus
\bigcup_{p\in E} b_p$.  For each point $p\in E$, let $h_p =
\partial\Sigma \cap \partial b_p$ be the concave surface of the
corresponding `hole' not eaten by any other ball, and let $H =
\bigcup_{p\in E} h_p$.  Equivalently, $H = \partial \Sigma \setminus
\partial D$.  See Figure~\ref{Fig:cheeseslice}.

We claim that the surface area of $H$ is only $O(r)$.  The proof of 
this claim is elementary but tedious; we give the proof separately 
below.  By an argument identical to Lemma 2.6 of \cite{e-npsch-01}, 
the ball of unit diameter centered at any point $p \in E$ contains at 
least $\Omega(1)$ of this surface area---for completeness, we also 
include this argument below.  Since these unit-diameter balls are 
disjoint, we conclude that $E$ contains at most $O(r)$ points.
\end{proof}

\begin{claim}
\label{c:cheese}
$\area(H) = O(r)$.
\end{claim}

\begin{proof}
Let $c_p$ be the common center of $B_p$ and $b_p$, and let $a_p$ be
the axis line through $c_p$ and normal to the planes
bounding~$\sigma$.  For any point $x\in h_p$, let $\bar{x}$ be its
nearest neighbor on the axis~$a_p$, and let $H_p$ be the union of
segments $x\bar{x}$ over all $x\in h_p$.  See Figure \ref{F:cheese}
for a two-dimensional example.  Finally, let $r_p = \min_{x\in h_p}
\abs{x \bar{x}}$.

The triangle inequality implies that $x\bar{x}$ and $y\bar{y}$ have
disjoint interiors whenever $x\ne y$, so
\begin{equation}
	\sum_{p\in E} \vol(H_p)
	=
	\vol \left(\bigcup_{p\in E} H_p\right)
	\le
	\vol(D) = O(r^2).
\label{eq:sumvol}
\end{equation}
We can bound the volume of each hole $H_p$ as follows:
\begin{align}
	\vol(H_p)
	&=
	\iint_{x\in h_p} \!\!
	\frac{\abs{x\bar{x}}\cdot \cos\angle c_p x \bar{x}}{2}\, dx^2
\notag
\\	&=
	\iint_{x\in h_p}\frac{\abs{x \bar{x}}^2}{2 \abs{x c_p}}\, dx^2
\notag
\\	&=
	\frac{1}{2 \abs{p c_p}}\, \iint_{x\in h_p} \abs{x \bar{x}}^2\, dx^2
\notag
\\	&\ge
	\frac{1}{12 r}\, \iint_{x\in h_p} \abs{x \bar{x}}^2\, dx^2
\notag
\\	&\ge
	\frac{r_p^2}{12 r}\area(h_p).
\label{eq:volHp}
\end{align}

\begin{figure}[htb]
\centerline{\includegraphics[height=2.35in]{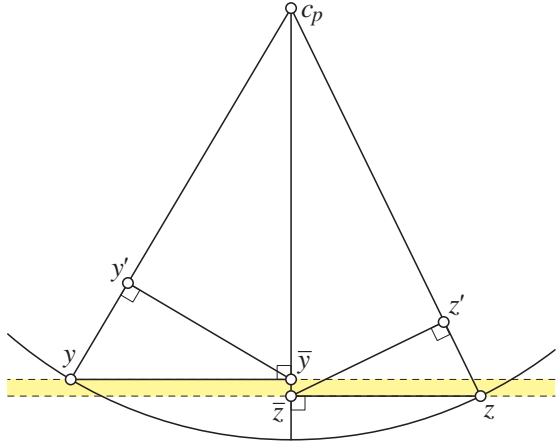}}
\caption{Proof of Claim \ref{c:cheese}.  The slab $\sigma'$ is shaded.}
\label{F:cheese}
\end{figure}
The intersection $b_p \cap \partial \sigma'$ consists of two parallel
disks, the smaller of which has radius $r_p$.  Since both $\sigma'$
and the boundary of $B_p$ contain the endpoints of some crossing edge
$pq$ of length at least~$2r$, the larger of these two disks has radius
at least $r - \omega - 4/3$.  Again referring to
Figure~\ref{F:cheese}, we choose two points $y, z\in h_p \cap
\partial\sigma' \subseteq b_p\cap \partial\sigma'$ on the boundary of
the larger and smaller disks, respectively, so that $r_p =
\abs{z\bar{z}}$.  We can bound this radius as follows:
\begin{align*}
	r_p^2
	=
	\Abs{z\bar{z}}^2
	&=
	\Abs{z c_p}^2 - \Abs{\bar{z} c_p}^2
\\	&=
	\Abs{z c_p}^2 - \left( \Abs{\bar{z}\bar{y}} + \Abs{\bar{y} c_p} \right)^2
\\	&=
	\Abs{z c_p}^2 -
	\left(	\Abs{\bar{z}\bar{y}} +
		\sqrt{\Abs{y c_p}^2 - \Abs{y\bar{y}}^2}
	\right)^2
\\	&=
	\Abs{z c_p}^2 -
	\left(	\Abs{\bar{z}\bar{y}}^2 +
		2 \Abs{\bar{z}\bar{y}}\sqrt{\Abs{y c_p}^2 - \Abs{y\bar{y}}^2} +
		\Abs{y c_p}^2 - \Abs{y\bar{y}}^2
	\right)
\\	&=
	\Abs{y\bar{y}}^2
	-
	2 \Abs{\bar{z}\bar{y}}\sqrt{\Abs{y c_p}^2 - \Abs{y\bar{y}}^2}
	-
	\Abs{\bar{z}\bar{y}}^2
\\	&\ge
	\Abs{y\bar{y}}^2
	-
	2 \Abs{\bar{z}\bar{y}}\Abs{y c_p}
	-
	\Abs{\bar{z}\bar{y}}^2.
\end{align*}
Now substituting the known equations and inequalities
\[
	\Abs{y\bar{y}} \ge r - \omega - 4/3, 	\qquad
	\Abs{\bar{z}\bar{y}} = \omega + 1,	\qquad
	\Abs{y c_p} < 4r - 1/3,
\]
we obtain the lower bound
\begin{equation}
	r_p^2
	\ge
	(r - 1/3)^2 - 2 (\omega+1) (6r - 1/3) - (\omega+1)^2
	=
	\Omega(r^2).
\label{eq:rp2}
\end{equation}
Finally, combining inequalities \eqref{eq:sumvol}, \eqref{eq:volHp},
and \eqref{eq:rp2} yields an upper bound for the surface area of $H$:
\[
	\area(H) =   \sum_{p\in E} \area(h_p)
		 \le \sum_{p\in E} \frac{12r}{r_p^2}\vol(H_p)
		 =   O(1/r) \cdot \sum_{p\in E} \vol(H_p)
		 =   O(r).
\]
\aftermath
\end{proof}

\begin{claim}[Lemma 2.6 of \cite{e-npsch-01}]
\label{c:eat}
Let $U$ be any unit-diameter ball contained in $\Sigma$ whose center 
is distance $1/3$ from~$H$.  Then $U$ contains $\Omega(1)$ 
surface area of~$H$.
\end{claim}

\begin{proof}
\begin{figure}[htb]
\centerline{\includegraphics[height=1.3in]{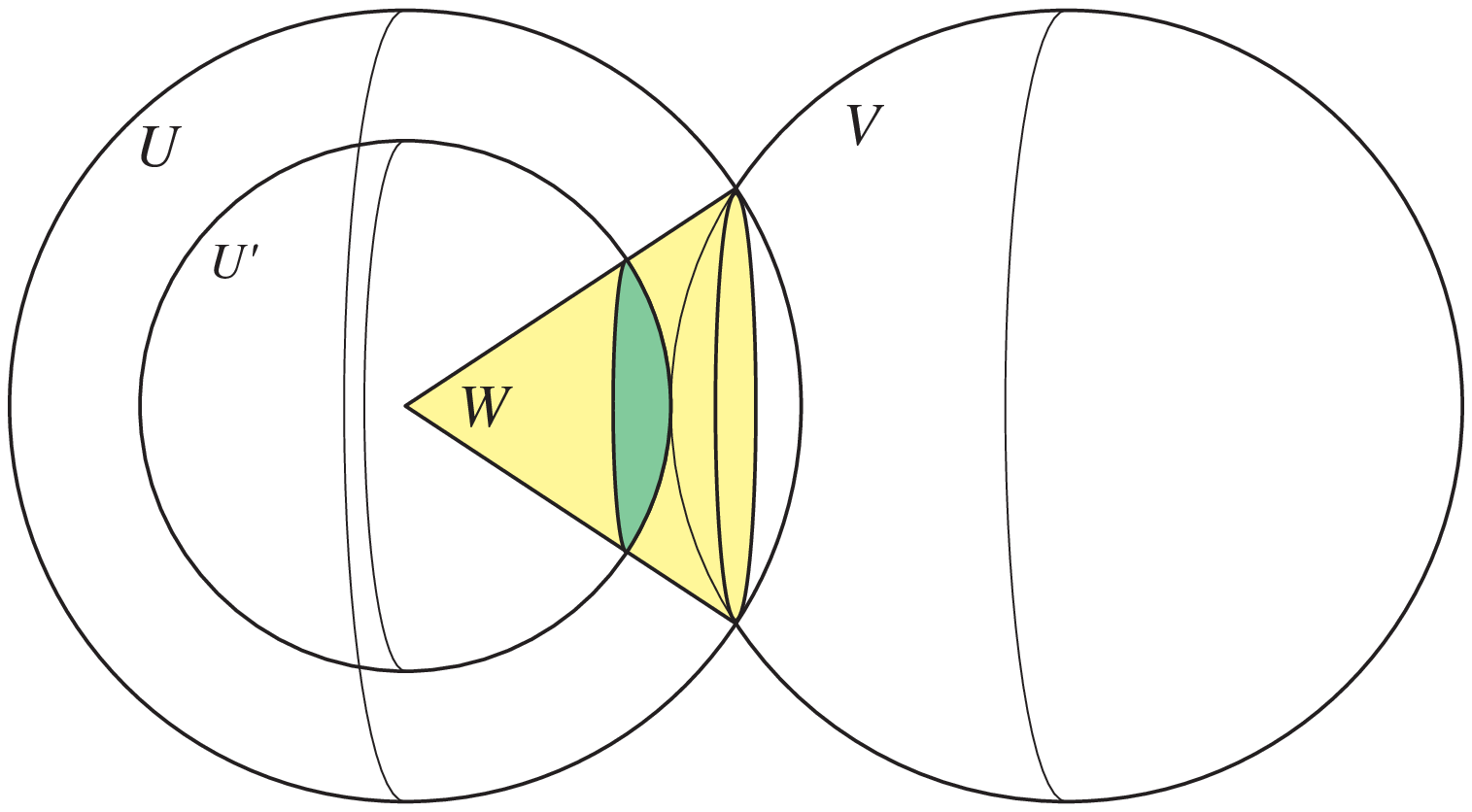}}
\caption{Proof of Claim \ref{c:eat}.}
\label{Fig/eat}
\end{figure}
Without loss of generality, assume that $U$ is centered at the origin
and that $(0, 0, 1/3)$ is the closest point of $H$ to the origin.  Let
$U'$ be the open ball of radius $1/3$ centered at the origin, let $V$
be the open unit ball centered at $(0, 0, 5/6)$, and let $W$ be the
cone whose apex is the origin and whose base is the circle $\partial
U\cap \partial V$.  See Figure~\ref{Fig/eat}.  $U'$ lies entirely
inside $\Sigma$, and since $r\ge 1$, we easily observe that $V$ lies
entirely outside $\Sigma$.  Thus, the surface area of $H \cap W
\subseteq H \cap U$ is at least the area of the spherical cap
$\partial U' \cap W$, which is $\pi/27$.
\end{proof}

Together, Lemmas \ref{L:slab}, \ref{L:degree}, and \ref{L:cheese}
imply that $O(r)$ relaxed edges intersect any pixel.  Since there are
$O(r^2)$ pixels, we conclude that there are $O(r^3)$ relaxed edges
overall.

\subsection{Tense Edges Are Easy to Relax}
\label{SS:twist}

In order to count the tense crossing edges of $P\cup Q$, we will show
that there are a constant number of transformations of space, such
that every tense edge is mapped to a relaxed edge at least once.

A \emph{\Mobius\ transformation} is a continuous bijection from the 
extended Euclidean space $\x{\Rd} = {\Rd \cup \set{\infty}} \simeq 
\Sphere^d$ to itself, such that the image of any sphere is a sphere.  
(A hyperplane in $\Rd$ is a sphere through $\infty$ in $\x{\Rd}$.)
The space of \Mobius\ transformations is generated by inversions.  
Examples include reflections (inversions by hyperplanes), translations 
(the composition of two parallel reflections), dilations (the 
composition of two concentric inversions), and the well-known 
stereographic lifting map from $\x{\Rd}$ to $\Sphere^d \subset 
\Real^{d+1}$ relating $d$-dimensional Delaunay triangulations to 
${(d+1)}$-dimensional convex hulls \cite{b-vdch-79}:
\[
	\lambda(x_1, x_2, \dots, x_d)
	=
	\frac{(x_1, x_2, \dots, x_d, 1)}
	     {x_1^2 + x_2^2 + \cdots + x_d^2 + 1},
	\qquad
	\lambda(\infty)
	=
	(0,0,\dots,0,0).
\]
\Mobius\ transformations are also the maps induced on the boundary of
hyperbolic space $\Hyper^{d+1}$ by hyperbolic isometries.
Two-dimensional \Mobius\ transformations are also called \emph{linear
fractional transformations}, since they can be written as maps on the
extended complex plane $\x{\Complex} = \Complex \cup \set\infty \simeq
\Sphere^2$ of the form $z \mapsto {(az+b)/(cz+d)}$ for some complex
numbers $a,b,c,d$.

\Mobius\ transformations are \emph{conformal}, meaning they locally
preserve angles.  There are infinitely many other two-dimensional
conformal maps~\cite{n-vca-99}---in fact, conformal maps are widely
used in algorithms for meshing planar domains \cite{dv-ncmuc-98} and
parameterizing surfaces \cite{eddhl-maam-95,lsscd-mmaps-98}---but
\Mobius\ transformations are the only conformal maps in dimensions
three and higher.  Higher-dimensional \Mobius\ transformations are
described in detail by Beardon \cite{b-gdg-83}; see also Hilbert and
Cohn-Vossen~\cite{hc-gi-52}, Thurston \cite{t-gt3m-97,t-tdgt1-97}, or
Miller \etal~\cite{mttv-sspnn-97}.

Let $\Sigma$ be a sphere in $\Real^3$ with finite radius (not passing
through the point $\infty$), and let $\pi:\x{\Real^3} \to \x{\Real^3}$
be a conformal transformation.  If $\pi(\Sigma)$ is also a
finite-radius sphere and the point $\pi(\infty)$ lies in the interior
of $\Pi(\Sigma)$, we say that $\pi$ \emph{everts} $\Sigma$.

Let $S$ be a set of points in $\Real^3 \subset \x{\Real^3}$, and let
$p,q,r,s\in S$ be the vertices of a Delaunay simplex with empty
circumsphere~$\Sigma$.  For any conformal transformation~$\kappa$, the
sphere $\kappa(\Sigma)$ passes through the points $\kappa(p)$,
$\kappa(q)$, $\kappa(r)$, and~$\kappa(s)$.  This sphere either
excludes every other point in $\kappa(S)$, contains every other point
in $\kappa(S)$, or is a plane with every other point of $\kappa(S)$ on
one side.  In other words,
$\conv\set{\kappa(p),\kappa(q),\kappa(r),\kappa(s)}$ is either a
Delaunay simplex, an anti-Delaunay\footnote{The anti-Delaunay
triangulation is the dual of the furthest point Voronoi diagram.}
simplex, or a convex hull facet of~$\kappa(S)$.  Thus, ignoring
degenerate cases, the abstract simplicial complex consisting of
Delaunay and anti-Delaunay simplices of any point set, which we call
its \emph{Delaunay polytope}, is invariant under conformal
transformations.

In this section, we exploit this conformal invariance to count tense
crossing edges.  The main idea is to find a small collection of
conformal maps, such that for any tense edge, at least one of the maps
transforms it into a relaxed edge, by shrinking (but not everting) its
circumsphere.  In order to apply our earlier arguments to count the
transformed edges, we consider only conformal maps that map $P\cup Q$
to another well-separated pair of sets with nearly the same spread.

\begin{figure}[htb]
\centerline{\includegraphics[height=2.5in]{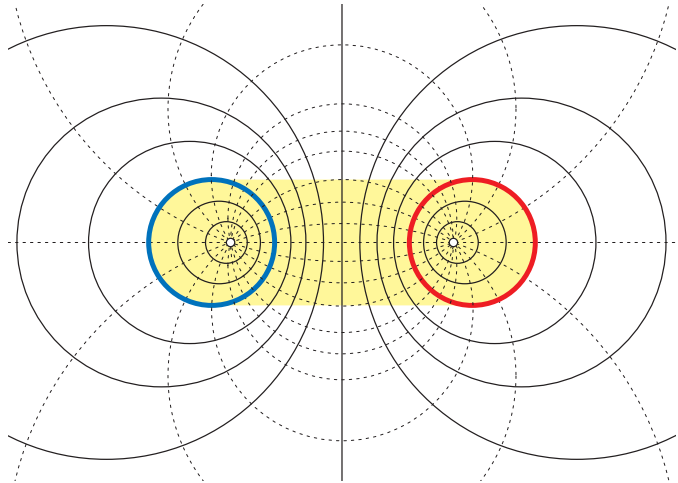}}
\caption{Every rotary map preserves every solid circle and maps every
dotted circle to another dotted circle.  The bold circles are
$\bigcirc P$ and $\bigcirc Q$.}
\label{Fig/conform}
\end{figure}

\medskip Recall that $P$ and $Q$ lie inside balls of radius $r$ 
centered at $(0,0,2r)$ and $(0,0,-2r)$, respectively.  Call these 
balls $\bigcirc P$ and $\bigcirc Q$.  We call an 
orientation-preserving conformal map $\kappa$ a \emph{rotary map} if 
it preserves these spheres, that is, if $\kappa(\bigcirc P) = \bigcirc 
P$ and ${\kappa(\bigcirc Q) = \bigcirc Q}$.  Rotary maps actually 
preserve a continuous one-parameter family of spheres centered on the 
$z$-axis, including the points $p^* = (0,0,\sqrt{3}r)$ and $q^* = 
(0,0,-\sqrt{3}r)$ and the plane $z=0$.  (In the space of spheres 
\cite{dmt-ssgtu-92t, e-dssd-99}, this family is just the line through 
$\bigcirc P$ and $\bigcirc Q$.)  See Figure~\ref{Fig/conform} for a
two-dimensional example.

The image of $P\cup Q$ under any rotary map is clearly well-separated.
In order to apply our earlier arguments, we also require that these
maps do not significantly change the spread.

\begin{lemma}
\label{L:stretch}
For any rotary map $\kappa$, the closest pair of points in
$\kappa(P\cup Q)$ has distance between $1/3$ and $3$.
\end{lemma}

\begin{proof}
Consider the stereographic lifting map $\lambda: \x{\Real^3} \to
\Sphere^3$ that takes $p^*$ and $q^*$ to opposite poles of~$\Sphere^3$
and the plane $z=0$ to the equatorial sphere.  Any rotary map can be
written as $\lambda^{-1} \circ \rho \circ \lambda$, where $\rho$ is a
simple rotation about the axis $\lambda(p^*)\lambda(q^*)$.  (Thus, the
space of rotary maps is isomorphic to $SO(3)$, the group of rigid
motions of~$\Sphere^2$.)

To make the stereographic lifting map $\lambda$ concrete, we embed
$\x{\Real^3}$ and $\Sphere^3$ into $\Real^4$, as the hyperplane $x_4 =
\sqrt{3}r$ and the sphere of radius $\sqrt{3}r/2$ centered at
$(0,0,0,\sqrt{3}r/2)$, respectively.  Now $\lambda$ is an inversion
through the sphere of radius $\sqrt{3}r$ centered at the origin $o =
(0,0,0,0)$:
\[
	\lambda(x_1, x_2, x_3, x_4)
		=
		\frac{3r^2 \big(x_1, x_2, x_3, x_4)}
		     {x_1^2 + x_2^2 + x_3^2 + x_4^2}.
	\qquad
	\lambda(\infty) = (0, 0, 0, 0)
\]
Simple calculations (see Beardon \cite[pp. 26--27]{b-gdg-83}) imply
that for any points $p,q \in \x{\Real^4}$, we have
\[
	\abs{\lambda(p)\,\lambda(q)}
	=
	\frac{3 r^2\, \abs{pq}}{\abs{po}\,\abs{qo}}.
\]
The distance from the origin $o$ to any point in $P\cup Q$ (in the
hyperplane $w=\sqrt{3}r$) is between $2r$ and $2\sqrt{3}r$.  Thus,
for any points $p,q \in P\cup Q$, we have
\[
	\frac{\abs{pq}}{4}
	\le 
	\abs{\lambda(p)\,\lambda(q)}
	\le 
	\frac{3\abs{pq}}{4}.
\]
Simple rotations do not change distances at all.  Thus, for any rotary
map $\pi$, we have
\[
	\frac{\abs{pq}}{3}
	\le 
	\abs{\pi(p)\,\pi(q)}
	\le 
	3\abs{pq}.
\]
for all points $p,q\in P\cup Q$.  The lemma follows immediately.
\end{proof}

\begin{lemma} 
\label{L:rotary}
There is a set of $O(1)$ rotary maps $\set{\pi_1, \pi_2, \dots, 
\pi_k}$ such that any crossing edge of $P\cup Q$ is mapped to a 
relaxed crossing edge of $\pi_i(P\cup Q)$ by some $\pi_i$.
\end{lemma}

\begin{proof}
Rotations about the $z$-axis are rotary maps, but since they do not
actually change the radius of any sphere, we would like to ignore
them.  We say that two rotary maps $\kappa_1$ and $\kappa_2$ are
\emph{rotationally equivalent} if $\kappa_1 = \rho \circ \kappa_2$ for
some rotation~$\rho$ about the $z$-axis.  The \emph{rotation class} of
a rotary map $\kappa$, which we denote~$\seq{\kappa}$, is the set of
maps that are rotationally equivalent to $\kappa$.  Since any rotation
class $\seq{\kappa}$ is uniquely identified by the point
$\kappa^{-1}(0,0,0)$ in the $xy$-plane, the space of rotation classes
is isomorphic to $\x{\Real^2} \simeq \Sphere^2$.

Let $B_1, B_2, \dots, B_m$ be the smallest empty balls containing the
crossing edges of $P\cup Q$.  For each ball~$B_i$, let~$\kappa_i$
denote any rotary map such that $\kappa_i(B_i)$ is centered on the
$z$-axis and is not everted, so $\kappa_i(B_i)$ is an empty Delaunay
ball of some crossing edge of $\kappa_i(P\cup Q)$.  We easily observe
that $\kappa_i(B_i)$ has radius less than $3r$, so the corresponding
crossing edge is relaxed.  Thus, for each crossing edge, we have a
point $\seq{\kappa_i}$ on the sphere of rotation classes,
corresponding to a rotation class of maps that relax that edge.

Our key observation is that we have a lot of `wiggle room' in choosing
our relaxing maps~$\kappa_i$.  Consider the ball $\bar{B}$ of radius
$3r$ centered at the origin; this is the smallest ball containing both
$\bigcirc P$ and $\bigcirc Q$.  Let $W$ be the set of rotation classes
$\seq{w}$ such that the radius of $w(\bar{B})$ is at most~$4r$ and
$w(\bar{B})$ is not everted.  $W$ is a circular cap of some constant
angular radius $\theta$ on the sphere of rotation classes, centered at
$\seq{1}$, the rotation class of the identity map.

For each $i$, the ball $\kappa_i(B_i)$ lies entirely inside $\bar{B}$,
so any rotation class in $W$ transforms $\kappa_i(B_i)$ into another
ball of radius at most $4r$.  Thus, \emph{any} rotation class in the
set $W_i = {\Set{\seq{w\circ\kappa_i} \mid \seq{w}\in W}}$ relaxes the
$i$th crossing edge.  $W_i$ is a circular cap of angular
radius~$\theta$ on the sphere of rotation classes, centered at the
point $\seq{\kappa_i}$.

\begin{figure}[htb]
\centering\includegraphics[height=2.25in]{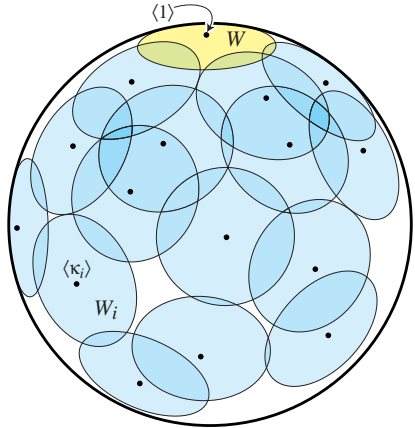}
\caption{For each crossing edge, there is a constant-radius cap on the
sphere of rotation classes.  A constant number of rotation classes
stab all these caps.}
\label{Fig:spherecaps}
\end{figure}

Since each of these $m$ caps has constant angular radius, we can stab
them all with a constant number of points.  Specifically, let $\Pi =
\set{\seq{\pi_1}, \seq{\pi_2}, \dots, \seq{\pi_k}} \subset \Sphere^2$
be a set of $k = O(1/\theta^2)$ points on the sphere of rotation
classes, such that any point in $\Sphere^2$ is within angular distance
$\theta$ of some point in $\Pi$.  (In surface reconstruction terms,
$\Pi$ is a $\theta$-sample of the sphere.)  Each disk $W_i$ contains
at least one point in~$\Pi$, which implies that each crossing edge is
relaxed by some rotation class $\seq{\pi_j}\in \Pi$.  Finally, to
satisfy the theorem, we choose an arbitrary rotary map~$\pi_j$ from
each rotation class ${\seq{\pi_j}\in \Pi}$.
\end{proof}

It follows immediately that $P\cup Q$ has $O(r^3)$ crossing edges.

\subsection{Charging Delaunay Edges to Volume}
\label{SS:split}

In the last step of our proof, we count the Delaunay edges in an
arbitrary point set $S$ by decomposing it into a collection of subset
pairs and counting the crossing edges for each pair.

Let $S$ be an arbitrary set of points with diameter~$\Delta$, where
the closest pair of points is at unit distance.  $S$ is contained in a
cube $\square S$ of width~$\Delta$.  We call an edge of the Delaunay
triangulation of $S$ \emph{short} if its length is less than $5$ and
\emph{long} otherwise.  A simple packing argument implies that $S$ has
at most $O(\Delta^3)$ short Delaunay edges.

To count the long Delaunay edges, we construct a \emph{well-separated
pair decomposition} of $S$~\cite{ck-dmpsa-95}, based on a simple
octtree decomposition of the bounding cube~$\square S$.  (See Agarwal
\etal~\cite{aesw-emstb-91} for a similar decomposition into subset
pairs.)  Our octtree has $\ceil{\log_2\Delta}$ levels.  At each
level~$\ell$, there are $8^\ell$ cells, each a cube of width $w_\ell =
\Delta/2^\ell$.  Our well-separated pair decomposition $\Xi =
\Set{(P_1, Q_1), (P_2, Q_2), \dots, (P_m, Q_m)}$ contains the points
in every pair of cells that are at the same level $\ell$ and are
separated by a distance between $3w_\ell$ and $6w_\ell$.

Every subset pair $(P_i, Q_i)\in \Xi$ is well-separated: if the pair
is at level $\ell$ in our decomposition, then for some $r_i =
\Theta(w_\ell)$, the sets $P_i$ and $Q_i$ lie in a pair of balls of
radius $r_i$ separated by distance~$2r_i$.  Thus, by our earlier
arguments, the Delaunay triangulation of $P_i \cup Q_i$ has at most
$O(w_\ell^3)$ crossing edges.

For any points $p,q\in S$ such that $\abs{pq}\ge 5$, there is a subset
pair $(P_i,Q_i) \in \Xi$ such that $p\in P_i$ and $q\in Q_i$.  In
particular, every long Delaunay edge of $S$ is a crossing edge between
some subset pair in~$\Xi$.  A straightforward counting argument
immediately implies that the total number of crossing edges, summed
over all subset pairs in $\Xi$, is $O(\Delta^3\log
\Delta)$~\cite{e-npsch-01}.  However, not every crossing edge appears
in the Delaunay triangulation of $S$.  We remove the final logarithmic
factor by charging crossing edges to volume as follows.

\begin{figure}[htb]
\centerline{\includegraphics[height=1.75in]{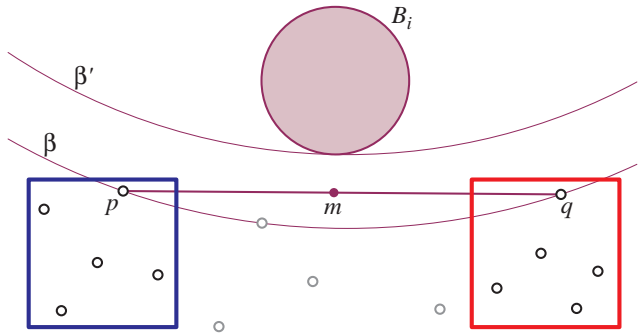}}
\caption{Defining the witness ball $B_i$ for a relevant subset pair.}
\label{F:charge}
\end{figure}

Say that a subset pair $(P_i, Q_i) \in \Xi$ is \emph{relevant} if some
pair of points $p_i\in P_i$ and $q_i\in Q_i$ are Delaunay neighbors
in~$S$.  For each relevant pair $(P_i, Q_i)$, we define a large,
close, and empty \emph{witness ball} $B_i$ as follows.  Choose an
arbitrary crossing edge $p_iq_i$ of $P_i\cup Q_i$.  If $P_i$ and $Q_i$
are at level $\ell$ in our decomposition, then the distance between
$p_i$ and~$q_i$ is at most ${(6+\sqrt{3})w_\ell}$.  Let~$\beta_i$ be
the smallest ball with $p_i$ and~$q_i$ on its boundary and no point of
$S$ in its interior; the radius of~$\beta_i$ is at least $3w_\ell/2$.
Let~$\beta'_i$ be a ball concentric with $\beta_i$ with radius smaller
by $w_\ell/2$.  Finally, let~$B_i$ be the ball of radius $w_\ell$
inside~$\beta'_i$ whose center is closest to the midpoint of segment
$p_iq_i$.  See Figure \ref{F:charge}.

$B_i$ is clearly empty.  The distance from any point in $B_i$ to any
point in $S$ is at least $w_\ell/2$, since $B_i \subset \beta'$.  On
the other hand, the triangle inequality implies that every point
in~$B_i$ has distance less than $(7+\sqrt{3})w_\ell/2 < 4.3661 w_\ell$
either to~$p$ or to~$q$, and thus to some cell at level $\ell$ in the
octtree.  It follows that if two witness balls overlap, their levels
differ by at most $\ceil{\lg (7+\sqrt{3})} = 4$.  Moreover, since any
ball of radius $(7+\sqrt{3})w_\ell/2$ intersects only a constant
number of cells at level $\ell$, at most a constant number of
level-$\ell$ witness balls overlap at any point.

For any relevant subset pair at level $\ell$, we can charge its
$O(w_\ell^3)$ crossing edges to its witness ball, which has volume
$\Omega(w_\ell^3)$.  Thus, the total number of relevant crossing edges
is at most the sum of the volumes of all the witness balls.  Since the
witness balls have only constant overlap, the sum of their volumes is
only a constant factor larger than the volume of their union.
Finally, every witness ball fits inside a cube of width $8\Delta$
concentric with $\square S$.  It follows that $S$ has at most
$O(\Delta^3)$ long Delaunay edges.

\medskip
This completes the proof of Theorem \ref{Th:del3}.

\section{Extensions and Implications}
\label{S:imps}

\subsection{Generalizing Spread}
\label{SS:relaxed}

Unfortunately, the spread of a set of points is an extremely fragile
measure.  Adding a single point to a set can arbitrarily increase its
spread, either by being too close to another point, or by being far
away from all the other points.  However, intuitively, adding a few
points to a set does not drastically increase the complexity of its
Delaunay triangulation.  (In fact, adding points can make the Delaunay
triangulation considerably \emph{simpler} \cite{ceghss-sdash-90,
beg-pgmg-94}.)  Clearly, our results can tolerate a small number of
outliers in the point set---up to $O(\Delta^3/n)$, to be precise---but
this is not very satisfying.

We can obtain a stronger result by generalizing the notion of spread.
For any integer $k$, define the \emph{order-$k$ spread} of a set $S$
to be the ratio of the diameter of $S$ to the radius of the smallest
ball that contains $k$ points of $S$.

\begin{theorem}
\label{Th:delk3}
The Delaunay triangulation of any set of points in $\Real^3$ with
order-$k$ spread $\Delta_k$ has complexity $O(k^2\Delta_k^3)$.
\end{theorem}

\begin{proof}
The proof of Theorem \ref{Th:del3} needs little modification to prove
this result; in fact, the only required changes are in the proofs of
Lemmas \ref{L:degree} and \ref{L:cheese}.

Let $P$ and $Q$ be two sets of points contained in balls of radius $r$
separated by distance $2r$, where any unit ball contains at most $k$
points.  As before, we separate $P$ and $Q$ with a grid of $O(r^2)$
circular pixels of constant radius $\e$.  Lemma \ref{L:slab} applies
verbatim.

A simple modification of the proof of Lemma~\ref{L:degree} implies
that if $\e<1/16$, then each point of $P$ is the endpoint of at most
$k$ relaxed edges through each pixel $\pi$.  Specifically, if $p$ is
the endpoint of $k+1$ edges $pq_0$, $pq_1, \dots, pq_k$ that all
intersect $\pi$, then some pair of points $q_i$ and $q_j$ would be
more than distance $1$ apart, which implies that either $pq_i$ or
$pq_j$ is not relaxed.  Similarly, since at most $k$ unit-diameter
balls overlap at any point, the set of relaxed endpoints $E\subset P$
contains at most $O(kr)$ points; the rest of the proof of
Lemma~\ref{L:cheese} in unchanged.

It follows that at most $O(k^2r)$ relaxed edges intersect any pixel,
so there are $O(k^2r^3)$ relaxed crossing edges overall.  Lemmas
\ref{L:stretch} and \ref{L:rotary} now imply that there are
$O(k^2r^3)$ crossing edges, and the well-separated pair decomposition
argument in Section \ref{SS:split} completes the proof.
\end{proof}

This generalization immediately implies the following high-probability
bound for random points.  Surprisingly, this result appears to be new.

\begin{corollary}
\label{C:random}
Let $S$ be a set of $n$ points generated independently and uniformly
in a cube in $\Real^3$.  The Delaunay triangulation of $S$ has
complexity $O(n\log n)$ with high probability.
\end{corollary}

\begin{proof}
Let $C$ be a cube of width $(n/\ln n)^{1/3}$.  If we generate $S$
uniformly at random inside $C$, the expected number of points in any
unit cube inside $C$ is exactly $\ln n$, and Chernoff's
inequality~\cite{mr-ra-95} implies that every unit cube inside $C$
contains $O(\log n)$ points with high probability.  Thus, with high
probability, $\Delta_k = O((n/\ln n)^{1/3})$ for some $k=O(\log n)$.
The result now follows immediately from Theorem \ref{Th:delk3}.
\end{proof}

\subsection{Unions of Several Dense Sets}
\label{SS:union}

We can also generalize our upper bound to sets that do not have small
spread, provided they can be decomposed into few subsets, where each
subset has low spread.  If all the subsets have the same `scale', the
upper bound is almost immediate. 

\begin{theorem}
\label{Th:weakunion}
Let $P_1, P_2, \dots, P_k$ be sets of points in $\Real^3$, each with
closest pair distance at least~$1$ and diameter at most~$\Delta$.  The
Delaunay triangulation of $P_1\cup P_2 \cup \cdots \cup P_k$ has
complexity $O(k^2\Delta^3)$.
\end{theorem}

\begin{proof}
It suffices to consider the case $k=2$; for larger values of $k$, we
separately count Delaunay edges for all $\binom{k}{2}$ pairwise unions.

Let $P$ and $Q$ be two sets, each with closest pair distance at least
$1$ and diameter at most $\Delta$.  We say that an edge in the
Delaunay triangulation off $P\cup Q$ is \emph{bichromatic} if it joins
a point in $P$ to a point in $Q$, and \emph{monochromatic} otherwise.
Theorem \ref{Th:del3} immediately implies that there are $O(\Delta^3)$
monochromatic edges.

If $P$ and $Q$ are well-separated, every bichromatic edge is a
crossing edge, so by our earlier analysis, there are $O(\Delta^3)$
bichromatic edges.  Otherwise, we can define a well-separated pair
decomposition by building an octtree over the bounding box of $P\cup
Q$, which has width at most $4\Delta$, so that every bichromatic edge
is a crossing edge for some well-separated subset pair.  By charging
bichromatic edges to empty witness balls exactly as before, we
conclude that the number of bichromatic edges is still $O(\Delta^3)$.
\end{proof}

This also provides an alternate proof of Theorem~\ref{Th:delk3}, since
any point set whose order-$k$ spread is~$\Delta_k$ can be partitioned
into $O(k)$ subsets satisfying the conditions of the theorem.

With more effort, we can establish a similar upper bound for unions of
arbitrary low-spread sets with arbitrarily different scales.  As
usual, we start by considering the case of two sets contained in
disjoint balls.  Let $P$ be a set of points with closest pair distance
$1$ inside a ball $\bigcirc P$ of radius $r$, and let $Q$ be a set of
points with closest pair distance $\delta\gg 1$ inside a ball
$\bigcirc Q$ of radius $R$.  We say that $P$ and $Q$ are
\emph{well-separated} if the distance between $\bigcirc P$ and
$\bigcirc Q$ is at least $r+R$.

First suppose $P$ and $Q$ are well-separated.  To analyze the number 
of crossing edges, we follow precisely the same outline as our earlier 
proof.  After an appropriate rigid motion, $\bigcirc P$ is centered at 
$(0,0,2r)$ and that $\bigcirc Q$ is centered at $(0,0,-2R)$.  We place 
a grid of $O(r^2)$ circular pixels of radius $\e = O(1)$ on the plane 
$z=0$, so that every crossing edge passes through a pixel.  The proof 
of Lemma \ref{L:slab} immediately implies that the crossing edges 
passing through any pixel lie in a slab of width $O(R/r)$ between two 
parallel planes.

Say that an crossing edge is \emph{relaxed} if its endpoints lie on an
empty sphere of radius $O(R+r)$.  The proof of Lemma \ref{L:degree}
immediately implies that each point in $Q$ is an endpoint of at most
one relaxed edge passing through any pixel.  (However, a point in $P$
might be an endpoint of more than one relaxed edge if $R/\delta < r$.)
Lemma \ref{L:cheese} generalizes as follows.

\begin{lemma}
\label{L:cheeseR}
The relaxed edges passing through any pixel $\pi$ are incident to at 
most $O(R/\delta + R^2/r\delta^2)$ points in $Q$.
\end{lemma}

\begin{proof}
Let $\sigma$ be the slab of width $O(R/r)$ containing the crossing
edges through $\pi$, and let $\sigma'$ be a parallel slab of width
$O(R/r + \delta)$ with the same central plane.  We define the `Swiss
cheese holes' $H$ exactly as in the proof of Lemma \ref{L:cheese}.
For each relaxed endpoint $q\in Q$, let $U_q$ be a ball of radius
$\delta/2$ centered at $q$; these balls are disjoint.  After an
appropriate scaling, Claim \ref{c:cheese} implies that the surface
area of $H$ is $O((R+r)(R/r + \delta)) = O(R^2/r + R\delta)$, and
Claim \ref{c:eat} implies that each ball~$U_q$ contains
$\Omega(\delta^2)$ of this surface area.
\end{proof}

Since there are $O(r^2)$ pixels, there are $O(r^2R/\delta + 
rR^2/\delta^2)$ relaxed crossing edges between $P$ and $Q$.  
Lemmas~\ref{L:stretch} and \ref{L:rotary} hold without modification, 
so the total number of crossing edges is $O(r^2R/\delta + 
rR^2/\delta^2)$.

Now suppose $P$ and $Q$ are \emph{not} well-separated.  We want to
count the crossing edges---Delaunay edges with one endpoint in each
set.  Let $\square P$ be a cube of width $r$ containing $P$, let $C$
be a concentric cube of width $8r$.  Say that a crossing edge is
\emph{short} if both endpoints are in $C$ and \emph{long} otherwise.
Since the spread of $Q\cap C$ is $O(r/\delta) = O(r)$, our earlier
well-separated pair decomposition argument implies that there are
$O(r^3)$ short crossing edges.

\begin{figure}[htb]
\centering\includegraphics[height=2in]{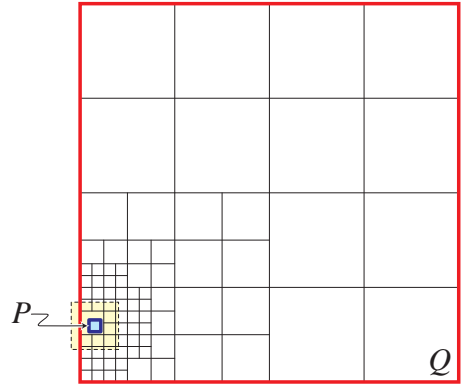}
\caption{Counting long crossing edges when $P$ and $Q$ are not
well-separated.}
\label{Fig:longcount}
\end{figure}

Let $\square Q$ be a cube of width $R$ containing $Q$.  To count the
long crossing edges, we construct an octtree over $\square Q$,
subdividing any cell that does not lie entirely inside $C$, whose
width is greater than $r$, and whose distance to $\square P$ is less
than its width plus $r$.  See Figure \ref{Fig:longcount}.  Let
$Q_\ell$ be the subset of $Q\setminus C$ inside a leaf cell $\ell$.
If the width of $\ell$ is $R/2^i$, then the spread of $Q_\ell$ is at
most $R/2^i\delta$.  If $Q_\ell$ is non-empty, then $P$ and $Q_\ell$
are well-separated.

By our earlier analysis, there are $O(r^2R/2^i\delta +
rR^2/4^i\delta^2)$ crossing edges between $P$ and $Q_\ell$.  Since
there are a constant number of leaf cells of any particular width,
there are less than
\[
	\sum_{i=1}^\infty
	O\!\left(
		\frac{r^2R}{2^i\delta} + \frac{rR^2}{4^i\delta^2}
	\right)
	=
	O\!\left(\frac{r^2R}{\delta} + \frac{rR^2}{\delta^2}\right)
\]
long crossing edges between $P$ and $Q\setminus C$.  Thus, the total 
number of crossing edges is $O(r^3 + r^2R/\delta + rR^2/\delta^2)$.

The spread of $P$ is $O(r)$, and the spread of $Q$ is $O(R/\delta)$.
If $r \le \Delta$ and $R/\delta \le \Delta$, then our upper bound on
the number of crossing edges simplifies to $O(\Delta^3)$.  Theorem
\ref{Th:del3} implies that there are also $O(\Delta^3)$ non-crossing
Delaunay edges, so the overall complexity of the Delaunay
triangulation of $P\cup Q$ is $O(\Delta^3)$.

Generalizing this analysis to more than two sets is straightforward.

\begin{theorem}
\label{Th:union}
Let $P_1, P_2, \dots, P_k$ be sets of points in $\Real^3$, each with
spread $\Delta$.  The Delaunay triangulation of $P_1\cup P_2 \cup
\cdots \cup P_k$ has complexity $O(k^2\Delta^3)$.
\end{theorem}

\subsection{Regular Triangulations}
\label{SS:regular}

Regular triangulations are perhaps the most natural generalization of
Delaunay triangulations.  Let $\hat{p} = (p, r(p))$ denote the ball
centered at point $p$ with radius $r(p)$; we can also think of
$\hat{p}$ as a point~$p$ with an associated weight $r(p)$.  Let
$\hat{S}$ be a set of balls (or equivalently, a set of weighted
points), and let $S$ be the set of centers of balls in $\hat{S}$.  The
\emph{power} from a point $x$ to a ball $\hat{p}\in \hat{S}$ is
$\abs{xp}^2 - r^2(p)$.  The \emph{power diagram} of $\hat{S}$ is the
Voronoi diagram with respect to this distance function.  The dual of
the power diagram is called the \emph{regular triangulation} of
$\hat{S}$.  The vertices of this triangulation are all points in $S$;
however, some points may not be vertices, as the corresponding region
in the power diagram is empty.  Regular triangulations can be
equivalently defined as the orthogonal (or stereographic) projection
of the lower convex hull of a set of points in one higher
dimension~\cite{e-gtmg-01}.

The empty circumsphere criterion for Delaunay triangulations
generalizes as follows.  We say that two balls $\hat{p}$ and $\hat{q}$
are \emph{orthogonal} if $\abs{pq} = r^2(p) + r^2(q)$, and
\emph{further than orthogonal} if $\abs{pq} > r^2(p) + r^2(q)$.  Any
sphere that is orthogonal to a set of balls is called an
\emph{orthosphere} of that set.  A subset of balls in $\hat{S}$ form a
simplex in the regular triangulation of $\hat{S}$ if it has an
\emph{empty} orthosphere, that is, an orthosphere that is further than
orthogonal from every other ball in $\hat{S}$.

\begin{theorem}
\label{Th:power}
The regular triangulation of any set of disjoint balls in $\Real^3$ 
whose centers have spread~$\Delta$ has complexity $O(\Delta^3)$.
\end{theorem}

\begin{proof}
Let $\hat{S}$ be a set of pairwise-disjoint balls, where the minimum
distance between any two centers is $1$, and the maximum distance
between any two centers is $\Delta$.  Note that the largest ball in
$\hat{S}$ has radius less than $\Delta$, which implies that $\hat{S}$
lies inside a ball of radius $2\Delta$.

Say that an edge $pq$ in the regular triangulation of $\hat{S}$ is
\emph{local} if $\abs{pq} < 8\min\set{r(p), r(q)}$.  We can charge
each local edge to whichever endpoint has larger radius.  By a
straightforward packing argument, each ball $\hat{p}\in\hat{S}$ is
charged at most $O(r(p)^3)$ times.  The volume of each ball~$\hat{p}$
is $\Omega(r(p)^3)$.  Since the balls are disjoint, the number of
local edges bounded by the total volume of the balls, which is
$O(\Delta^3)$.

To count non-local edges, we follow the same outline as the proof of
Theorem \ref{Th:del3}.  First consider the well-separated case.  Let
$\hat{P}$ and $\hat{Q}$ be two sets of balls whose centers lie in
balls of radius $r$ separated by distance $2r$.  We claim that the
regular triangulation of $\hat{P}\cup\hat{Q}$ has $O(r^3)$ non-local
crossing edges.  Without loss of generality, we can assume that every
ball in $\hat{P}\cup\hat{Q}$ has radius at most $r/4$, since any
larger ball has only local crossing edges.  Thus, the empty
orthospheres of any crossing edge have radius larger than $3r/4$, and
the bounding spheres of $\hat{P}$ and $\hat{Q}$ have radius at most
$5r/4$ and are separated by distance at least $3r/2$.  We modify the
proof of Theorem \ref{Th:del3} by using smallest empty orthospheres
instead of empty circumspheres.  This replacement increases the
constants, but the proofs of most of the lemmas require no other
modification.  We will describe only the necessary modifications here;
refer to the earlier proofs for notation and definitions.

Edelsbrunner actually proved that \emph{regular} triangulations have 
a consistent depth order from any viewpoint \cite{e-atccd-90, 
e-gtmg-01}.  Thus, Lemma \ref{L:screw} holds with no modification.

The proof of Lemma \ref{L:cheese} requires one qualitative change.
First, we shrink each orthosphere $B_p$ by only $1/8$ (instead of
$1/3$) to obtain $b_p$; this does not substantially affect the proof
of Claim~\ref{c:cheese}.  We define a new ball $U_p$ around each
endpoint $p\in E$ as follows.  If $r(p)\le 1/2$, then~$U_p$ is the
unit-diameter ball centered at $p$.  Otherwise, $U_p$ is any ball of
radius $1/4$ inside~$\hat{p}$ whose center lies on $B_p$; such a ball
always exists if $r>10$.  A simple modification of the proof of Claim
\ref{c:eat} implies that each ball $U_p$ contains $\Omega(1)$ surface
area of the Swiss cheese slice~$\Sigma$.  It follows that
$\hat{P}\cup\hat{Q}$ has $O(r^3)$ relaxed non-local crossing edges.

The abstract complex consisting of the regular and
anti-regular\footnote{dual to vertices of the furthest-ball power
diagram} simplices of $\hat{P}\cup\hat{Q}$ is invariant under almost
any conformal transformation.  We can easily adapt the proof of Lemma
\ref{L:stretch} to show that for any rotary transformation $\pi$, the
spread of the centers of $\pi(\hat{P})$ is at most a constant factor
larger than the spread of the centers of $\hat{P}$.  (Note that the
center of the transformed sphere $\pi(\hat{p})$ is not necessarily the
image $\pi(p)$ of the original center.)  Thus, every rotary image
$\pi(\hat{P}\cup\hat{Q})$ has $O(r^3)$ relaxed non-local crossing
edges.  Lemma \ref{L:rotary} requires no modification, so
$\hat{P}\cup\hat{Q}$ has $O(r^3)$ non-local crossing edges.

Finally, we slightly modify our well-separated pair decomposition
argument, by defining a subset pair $(\hat{P}_i, \hat{Q}_i)$ to be
relevant if and only if it contributes a \emph{non-local} crossing
edge to the regular triangulation.  If this is the case, we define
$\beta_i$ to be the smallest empty orthosphere for any such non-local
edge.  The remainder of the argument is unchanged.  We conclude that
the regular triangulation of $\hat{S}$ has $O(\Delta^3)$ non-local
edges.
\end{proof}

We can generalize this upper bound further by allowing the balls to
overlap slightly.  A set of balls forms a \emph{$k$-ply system} if no
point in space is covered by more than $k$ balls \cite{mttv-sspnn-97}.
Applying precisely the same modifications as in the proof of
Theorem~\ref{Th:delk3}, we obtain the following result.

\begin{theorem}
\label{Th:kply}
The regular triangulation of any $k$-ply system of balls in $\Real^3$
whose centers have spread $\Delta$ has complexity $O(k^2\Delta^3)$.
\end{theorem}

A special case of a $k$-ply system is the \emph{hard sphere model}
commonly used in molecular modeling~\cite{m-ms-90}.  A set of balls is
\emph{hard} if the largest and smallest radii differ by a constant
factor $r$ and, after shrinking each ball by a constant factor $\rho$,
no ball contains the center of any other ball.  Halperin and Overmars
\cite{ho-smhsr-98} proved that any hard set of balls forms a $k$-ply
system of balls, where $k = O(r^3\rho^3) = O(1)$.  (See also Halperin
and Shelton \cite{hs-pssaa-98}.)

\begin{corollary}
\label{C:hard}
The regular triangulation of any hard set of balls in $\Real^3$ whose
centers have spread~$\Delta$ has complexity $O(\Delta^3)$.
\end{corollary}

Combining the ideas in Theorems \ref{Th:delk3}, \ref{Th:union}, and
\ref{Th:kply}, we obtain similar bounds for any sets of balls that can
be partitioned into a constant number of subsets, each of which is a
constant-ply system whose centers have small constant-order spread.
We omit further details.

\subsection{Surface Data}
\label{SS:surface}

A somewhat less obvious implication concerns dense surface data.  Let
$\Sigma$ be a $C^2$ surface in $\Real^3$.  Recall from Section
\ref{SS:oldsurf} that a point set $S$ is a \emph{uniform $\e$-sample}
of $\Sigma$ if, for some constant ${0<\delta<1/2}$, the distance
between any surface point $x\in \Sigma$ to the second-closest sample
point in $S$ is between $\delta\e\lfs(x)$ and $\e\lfs(x)$.

\begin{theorem}
\label{Th:surf}
Let $\Sigma$ be a fixed $C^2$ surface in $\Real^3$.  The Delaunay
triangulation of any uniform $\e$-sample of $\Sigma$ has complexity
$O(n^{3/2})$.
\end{theorem}

\begin{proof}
Let $S$ be a uniform $\e$-sample of $\Sigma$.  This set contains $n = 
\Theta(\mu/\e^2)$ points, where $\mu$ is the \emph{sample measure} 
of~$\Sigma$ \cite[Lemma 3.1]{e-npsch-01}.  The spread of $S$ is 
$\Theta(\Delta/\e)$, where $\Delta$ is the \emph{spread} of $\Sigma$, 
the ratio between the diameter of $\Sigma$ and its minimum local 
feature size.  Thus, by Theorem~\ref{Th:del3}, the Delaunay 
triangulation of $S$ has complexity $O(\Delta^3/\e^3) = O(n^{3/2} 
\Delta^3/\mu^{3/2}) = O(n^{3/2})$.
\end{proof}

This bound is tight in the worst case, for example, when $\Sigma$ is a
circular cylinder with spherical caps~\cite{e-npsch-01}.  Note that
Theorem \ref{Th:surf}, which applies to any fixed surface, does
\emph{not} contradict our earlier $\Omega(n^2/\log^2 n)$ lower bound,
which requires the surface to depend on $n$ and~$\e$.

Attali and Boissonnat \cite{ab-cdtpp-01} recently showed that under
certain sampling conditions, samples of \emph{polyhedral} surfaces
have linear-complexity Delaunay triangulations, improving earlier
subquadratic bounds \cite{ab-cdtpp-01}.  Unlike most
surface-reconstruction results, their sampling conditions do not take
local feature size into account (since otherwise samples would be
infinite).  They define a point set $S$ to be a $(\e, \kappa)$-sample
of $\Sigma$ if the ball of radius $\e$ centered at any surface point
contains at least $1$ and at most $\kappa$ points in $S$; they then
show that the Delaunay triangulation of any $(\e, \kappa)$-sample of a
fixed polyhedral surface has complexity $O(n)$.

\begin{theorem}
\label{Th:surf2}
Let $\Sigma$ be a fixed (not necessarily polyhedral or smooth) surface
in $\Real^3$.  The Delaunay triangulation of any $(\e, \kappa)$-sample
of $\Sigma$ has complexity $O(\kappa^2 n^{3/2})$, where $n$ is the
number of sample points.
\end{theorem}

\begin{proof}
Let $S$ be an $(\e, \kappa)$-sample of $\Sigma$.  This set contains $n
= \Omega(1/\e^2)$ points, where the hidden constant is proportional to
the surface area of $\Sigma$.  The order-$\kappa$ spread of $S$ is
$O(1/\e) = O(\sqrt{n})$.  The result now follows immediately from
Theorem~\ref{Th:delk3}.
\end{proof}

Again, this bound is tight (at least for constant $\kappa$) in the
case of a cylinder.  Attali and Boissonnat conjecture that for any
\emph{generic} fixed surface, the Delaunay triangulation of any
$(\e,\kappa)$-sample has near-linear complexity; they define a surface
to be generic if every medial ball meets the surface in a constant
number of points.

Finally, we consider the case of randomly distributed points on 
surfaces.

\begin{theorem}
\label{Th:randsurf}
Let $\Sigma$ be a fixed surface with surface area $1$, and let $S$ be 
a set of points generated by a homogeneous Poisson process on $\Sigma$ 
with rate $n$.  With high probability, the Delaunay triangulation of 
$S$ has complexity $O(n^{3/2}\log^{1/2} n)$.
\end{theorem}

\begin{proof}
let $\sigma$ be a ball of radius $O(\sqrt{(\log n)/n})$ centered on
$\Sigma$.  The expected number of points in $S\cap\sigma$ is $O(\log
n)$, and Chernoff's inequality~\cite{mr-ra-95} implies that
$S\cap\sigma$ contains between $1$ and $O(\log n)$ points with high
probability.  We can cover $\Sigma$ with less than $n$ such balls.
Thus, $S$ is an $(\e, \kappa)$-sample with high probability, where
$\e=O(\sqrt{(\log n)/n})$ and $\kappa=O(\log n)$.  The result now
follows from Theorem~\ref{Th:surf2}.
\end{proof}

Unlike most of the other results in this paper, this upper bound is
almost certainly not tight.  We conjecture that for \emph{any} fixed
piecewise-smooth surface, the Delaunay triangulation of a sufficiently
dense uniform random sample has near-linear complexity with high
probability.  Recent experimental results of Choi and Amenta
\cite{ca-dtpsd-02} support this conjecture.

\subsection{Algorithms}
\label{SS:algo}

Finally, our upper bounds immediately imply that several existing
algorithms based on Delaunay triangulations are more efficient if the
input point set is dense.

\begin{theorem}
\label{Th:fast}
The Delaunay triangulation of any set of $n$ points in $\Real^3$ with
spread $\Delta$ can be computed in $O(\Delta^3\log n)$ expected time, 
or in $O(\Delta^3\log^2 n)$ worst-case time.
\end{theorem}

\begin{proof}
To obtain the expected time bound, we apply the standard randomized
incremental algorithm of Guibas, Knuth, and Sharir
\cite{gks-ricdv-92}, which inserts the points one at a time in random
order; see also \cite{es-itfwr-96, cs-arscg-89}.  The running time of
this algorithm can be broken down into a \emph{point location} and
\emph{repair} phases.  In the point location phase, the algorithm
locates the Delaunay simplex containing the next point.  The total
time for all the point location phases is $O(n\log n)$.  The repair
phase actually inserts the point, repairs the Delaunay triangulation,
and updates the point-location data structure.  If the newly inserted
point has degree $k$ in the updated Delaunay triangulation, then its
repair phase costs $O(k)$ time.

Inserting a point into a set can only increase its spread, by either
increasing the diameter, decreasing the closest pair distance, or
both.  Thus, at all stages of the algorithm, the spread of the points
inserted so far is at most $\Delta$, so every intermediate Delaunay
triangulation has complexity $O(\Delta^3)$.  It follows that the
expected degree of the $i$th inserted point, and thus the time for the
$i$th repair phase, is $O(\Delta^3/i)$.  Therefore, the total time for
all repair phases is $\sum_{i=1}^n O(\Delta^3/i) = O(\Delta^3\log n)$.
This dominates the total point-location time.

The worst-case bound follows immediately from the deterministic 
output-sensitive algorithm of Chan, Snoeyink, and 
Yap~\cite{csy-pddpo-97}.
\end{proof}

Since the Euclidean minimum spanning tree of a set of points is a
subcomplex of the Delaunay triangulation, we can also compute it in
$O(\Delta^3\log n)$ expected time, by first computing the Delaunay
triangulation and then running any efficient minimum spanning tree
algorithm on its $O(\Delta^3)$ edges.  This immediately improves the
$O(n^{4/3+\e})$-time algorithm of Agarwal \etal~\cite{aesw-emstb-91}
whenever $\Delta = O(n^{4/9})$.  We can similarly improve the running
times for computing other Delaunay substructures, such as Gabriel
complexes, $\alpha$-shapes \cite{eks-sspp-83}, wrap and flow complexes
\cite{e-srwfp-02,gj-fcdsg-03,gj-nddp-02,s-vdmtd-99}, and cocone
triangles~\cite{acdl-sahsr-00}.

\begin{theorem}
Any set of $n$ points in $\Real^3$ with spread $\Delta$ can be stored 
in a data structure of size $O(\Delta^3\log n)$, so that nearest 
neighbor queries can be answered in $O(\log^2 n)$ time.
\end{theorem}

\begin{proof}
We construct a bottom-vertex triangulation of the Voronoi diagram of
the points, using a standard randomized incremental algorithm, where
the history graph is used as a point-location data structure.  A
similar algorithm for building (radial triangulations of)
two-dimensional Voronoi diagrams is described by
Mulmuley~\cite[Chapter 3.3]{m-cgitr-93}.  The running time analysis is
almost identical to the proof of Theorem~\ref{Th:fast}.

To speed up the search time, we add auxiliary point-location data
structures to the history graph.  Each insertion destroys several
tetrahedra and creates new ones.  We cluster the new tetrahedra
according to which Voronoi cells contain them.  For each cluster, we
construct a point-location structure to determine, given a point $q$
inside that cluster, which tetrahedron contains~$q$.  Because each
cluster of tetrahedra shares a common vertex, we can use \emph{planar}
point-location structures with linear size and logarithmic query
time~\cite{as-ewcqc-00}.  With high probability, the simplex
containing any query point~$q$ changes $O(\log n)$ times during the
incremental construction of the Voronoi diagram.  When the simplex
containing $q$ changes, we can determine which cluster to query by a
simple distance comparison.  Thus, the total time to locate $q$ in the
final Voronoi diagram is $O(\log^2 n)$ with high probability.  The
total space used by the auxiliary structures is bounded by the size of
the history graph, which is $O(\Delta^3\log n)$ on average.  In
particular, some ordering of the points gives us a data structure of
size $O(\Delta^3\log n)$ with worst-case query time $O(\log^2 n)$.
\end{proof}


\section{Denser Regular Triangulations}
\label{S:nondel}

Despite our success in the previous two sections, our $O(\Delta^3)$
upper bound does not generalize to arbitrary triangulations, or even
arbitrary regular triangulations.  Recall that regular triangulations
in $\Real^3$ can be defined as the orthogonal projection of the lower
convex hull of a set of points in $\Real^4$.

\begin{theorem}
\label{Th:regular}
For any $n$ and $\Delta < n$, there is a set of $n$ points with spread
$\Delta$ with a regular triangulation of complexity $\Omega(n\Delta)$.
\end{theorem}

\begin{proof}
Any affine transformation of $\Real^3$ lifts to an essentially unique
affine transformation of $\Real^{d+1}$ that preserves vertical lines
and vertical distances.  Since affine transformations preserve
convexity, it follows that any affine transformation of a regular
triangulation is another regular triangulation (but possibly with
\emph{very} different weights).  Thus, to prove the theorem, it
suffices to construct a set $S$ of $n$ points whose \emph{Delaunay}
triangulation has complexity $\Omega(n\Delta)$, such that some affine
image of $S$ has spread $O(\Delta)$.

Without loss of generality, assume that $\sqrt{n/\Delta}$ is an
integer.  For each positive integer $i,j\le \sqrt{n/\Delta}$, let
$s(i,j)$ be the line segment with endpoints $(2i,2j,0) \pm
((-1)^{i+j}, (-1)^{i+j}, 1)$.  Let $S$ be the set of $n$ points
containing $\Delta$ evenly spaced points on each segment $s(i,j)$.
Straightforward calculations imply that the Delaunay triangulation of
$S$ contains at least $\Delta^2/4$ edges between any segment $s(i,j)$
and any adjacent segment ${s(i\pm 1, j)}$ or $s(i, j\pm 1)$.  Thus,
the overall complexity of the Delaunay triangulation of~$S$ is
$\Omega(n\Delta)$.  Applying the linear transformation $f(x,y,z) =
(x,y,\Delta z)$ results in a point set $f(S)$ with spread $O(\Delta)$.
\end{proof}

This result does not contradict Theorems \ref{Th:power} or
\ref{Th:kply}, since the weighted points in our construction are
equivalent to balls that overlap heavily.  In fact, the largest ball
in our construction actually contains the centers of a constant
fraction of the other balls.


\section{Open Problems}
\label{S:outro}

Our results suggest several open problems, the most obvious of which
is to simplify our rather complicated proof of Theorem~\ref{Th:del3}.
The hidden constant in our upper bound is in the millions; the
corresponding constant in the lower bound (which seems much closer to
the true worst-case complexity) is about~$10$.

We conjecture that Theorem \ref{Th:regular} is tight for 
\emph{arbitrary} triangulations.  In fact, we believe that any complex 
of points, edges, and triangles, embedded in $\Real^3$ so that no 
triangle crosses an edge, has $O(n\Delta)$ triangles.  Even the 
following special case is still open: What is the minimum spread of a 
set of $n$ points in $\Real^3$ in which \emph{every} pair is joined by 
a Delaunay edge?  We optimistically conjecture that the answer is 
$n/\pi - o(n)$; this is the spread of $n$ evenly-spaced points on a 
single turn of a helix with infinitesimal pitch.

What is the worst-case complexity of the convex hull of a set of $n$
points in $\Real^4$ with spread~$\Delta$?  Our earlier results
\cite{e-npsch-01} already imply a lower bound of
$\Omega(\min\set{\Delta^3, n\Delta, n^2})$.  This bound is \emph{not}
improved by Theorem \ref{Th:regular}, since our construction requires
points with extremely large weights.  The only known upper bound is
$O(n^2)$.

Another interesting open problem is to generalize the results in this
paper to higher dimensions.  Our techniques almost certainly imply an
upper bound of $O(\Delta^d)$ on the number of Delaunay \emph{edges},
improving our earlier upper bound of $O(\Delta^{d+1})$.
Unfortunately, this gives a very weak bound on the overall complexity,
which we conjecture to be $O(\Delta^d)$.  What is needed is a
technique to directly count $\floor{d/2}$-dimensional Delaunay
simplices: triangles in~$\Real^4$, tetrahedra in~$\Real^6$, and so on.

Standard range searching techniques can be used to answer nearest 
neighbor queries in~$\Real^3$ in $O(\log n)$ time using 
$O(n^2/\polylog n)$ space, or in $O(\sqrt{n}\polylog n)$ time using 
$O(n)$ space \cite{survey, c-chdc-93, c-racpq-88, tradeoff, 
m-rsehc-93}.  Using these data structures, we can compute the 
Euclidean spanning tree of a three-dimensional point set in 
$O(n^{4/3+\e})$ time \cite{aesw-emstb-91}.  All these results 
ultimately rely on the simple observation that the Delaunay 
triangulation of a random sample of a point set is significantly less 
complex (in expectation) than the Delaunay triangulation of the whole 
set.  Unfortunately, if we try to reanalyze these algorithms in terms 
of the spread, this argument falls apart---in the worst case, a random 
sample of a point set with spread $\Delta$ has expected spread close 
to $\Delta$, so the Delaunay triangulation of the subset is \emph{not} 
significantly simpler after all!  Can random sampling be integrated 
with our distance-sensitive bounds?  Is there a data structure of size 
$O(n)$ that supports faster nearest neighbor queries when the spread 
is, say, $O(\sqrt{n})$?

\paragraph{Acknowledgments.}
Once again, I thank Edgar Ramos for suggesting well-separated pair
decompositions.  Thanks also to Edgar Ramos, Herbert Edelsbrunner, Pat
Morin, Sariel Har-Peled, Sheng-Hua Teng, Tamal Dey, and Timothy Chan
for helpful discussions; to Pankaj Agarwal and Sariel Har-Peled for
comments on an earlier draft of the paper; and to the anonymous SODA
reviewers for pointing to the work of Miller \etal~\cite{mtg-ocum-99,
mttv-sspnn-97, mttw-dbnmt-95}.

\bibliographystyle{newabuser}
\bibliography{screw,jeffe,career,geom}


\end{document}